\documentclass[preprint,1p]{elsarticle}
\usepackage{amsmath,amssymb}
\usepackage{graphicx}
\usepackage{dcolumn}
\usepackage{bm}
\usepackage{epstopdf}
\usepackage{subfigure}
\usepackage{dsfont} 
\usepackage{natbib}
\newcommand{\ti}{\text{i}}
\usepackage[colorlinks=true,linkcolor=blue,urlcolor=blue,citecolor=blue]{hyperref}
\usepackage{makecell}
\usepackage{pgf}
\journal{Physica D}

\begin{document}

\title{Hydroelastic lumps in shallow water}
\author{Yanghan Meng}
\address{Institute of Mechanics, Chinese Academy of Sciences,Beijing 100190, China}
\address{School of Engineering Science, University of Chinese Academy of
	Sciences, Beijing 100049, China}
\author{Zhan Wang}%
\ead{zwang@imech.ac.cn}
\address{Institute of Mechanics, Chinese Academy of Sciences,Beijing 100190, China}
\address{School of Engineering Science, University of Chinese Academy of
 Sciences, Beijing 100049, China}
\address{School of Future Technology, University of Chinese Academy of
 Sciences, Beijing 100049, China}
%

\date{\today}

\begin{abstract}
Hydroelastic solitary waves propagating on the surface of a three-dimensional ideal fluid through the deformation of an elastic sheet are studied. The problem is investigated based on a Benney-Luke-type equation derived via an explicit non-local formulation of the classic water wave problem. The normal form analysis is carried out for the newly developed equation, which results in the Benney-Roskes-Davey-Stewartson (BRDS) system governing the coupled evolution of the envelope of a carrier wave and the wave-induced mean flow. Numerical results show three types of free solitary waves in the Benney-Luke-type equation: plane solitary wave, lump (i.e., fully localized traveling waves in three dimensions), and transversally periodic solitary wave, and they are counterparts of the BRDS solutions. They are linked together by a dimension-breaking bifurcation where plane solitary waves and lumps can be viewed as two limiting cases, and transversally periodic solitary waves serve as intermediate states. The stability and interaction of solitary waves are investigated via a numerical time integration of the Benney-Luke-type equation. For a localized load moving on the elastic sheet with a constant speed, it is found that there exists a transcritical regime of forcing speed for which there are no steady solutions. Instead, periodic shedding of lumps can be observed if the forcing moves at speed in this range.
\end{abstract}

\maketitle


\section{Introduction}
Vibrations of a thin elastic plate on top of a fluid body due to gravity, elasticity, and external load are known as hydroelastic waves or flexural-gravity (FG) waves. Hydroelastic waves have wide applications in the safe use of floating ice covers as transport links and the construction of very large floating platforms. Typical lengths of hydroelastic waves are generally from tens to hundreds of meters, and then these waves are easily observed in field campaigns. Two famous field studies on hydroelastic waves were conducted respectively in McMurdo Sound near Antarctica for the case of deep water \cite{Squire1988} and Lake Saroma in Japan for shallow water \cite{Takizawa1985}. Recently, \citet{VanderSanden2017} measured the ice deflection induced by the motion of a vehicle based on the satellite synthetic-aperture radar.

In the past few decades, many analytical and numerical studies of hydroelastic waves concentrated on the two-dimensional (2D) problem. The Euler-Bernoulli beam theory was extensively used in early works, which is a good approximation for small-amplitude deformations of the elastic sheet. However, this theory is not so suitable when the deformation becomes large. Therefore, the nonlinear Kirchhoff-Love (KL) plate theory was later developed for describing large-amplitude deformations. There have been rich studies based on this nonlinear elastic model. \citet{Parau2002} derived the forced nonlinear Schr\"{o}dinger (NLS) equation which predicted the existence of solitary waves if the fluid depth was in a specific range, and performed numerical computations of the full Euler equations for validation. \citet{Milewski2011} considered hydroelastic waves in deep water via asymptotic and numerical methods for both forced and unforced problems. They found a transcritical range of load speed where the shedding of solitary waves occurred for large external forces in unsteady simulations. It is noted that in deep water, hydroelastic solitary waves exist only at finite amplitude, which is coincident with the defocusing nature of the NLS equation at the phase speed minimum. \citet{Milewski20131} numerically found dark solitary waves in deep water in the full Euler equations using the conformal mapping technique, which were also predicted by the defocusing NLS equation. 

Recently, \citet{Toland2007} proposed a new model for ice sheets based on the Cosserat theory of hydroelastic shells satisfying Kirchhoff's hypotheses. The chief advantage of the Toland model over the KL model is that it conserves the elastic potential energy and hence has a clear variational structure. \citet{Guyenne2012} computed both depression and elevation solitary waves below the phase speed minimum in deep water based on this model. Their numerical results were complemented by \citet{Wang2013}, who found new highly nonlinear elevation solitary waves using a numerical continuation method. \citet{Gao2016} discovered a new class of non-symmetric solitary waves arising via spontaneous symmetry-breaking bifurcations. Periodic and generalized hydroelastic solitary waves were explored extensively by \citet{Gao2014} using a series truncation method. 

Linear models for both elasticity and fluid flow were used in early works of the three-dimensional (3D) problem, and hence the asymptotic Fourier analysis technique was applicable. Of note is the work of \citet{Davys1985}, who considered waves generated by a concentrated moving point load and obtained wave patterns, and \citet{Milinazzo1995}, who investigated the steady response of a uniform ice sheet induced by a rectangular moving load. Recently, \citet{Parau2011,Dinvay2019} investigated various types of responses of a floating ice sheet of infinite extent for disturbances moving at different velocities considering the nonlinear effect of the flow. \citet{Milewski20132} derived a Benney-Roskes-Davey-Stewartson (BRDS) system in the small-amplitude modulational asymptotic limit, on the base of which they studied the bifurcation mechanism of lumps together with their stability properties for a range of depths.  Free hydroelastic lumps were computed by \citet{Wang2014} using a quintic truncation model in the Hamiltonian framework and by \citet{Trichtchenko2018} based on a boundary integral equation method.

There are challenges to the 3D hydroelastic wave problem due to the complexity of the full Euler equations. Therefore, simplified equations in various asymptotic regimes are necessary for reducing model complexity and computational cost. \citet{Guyenne2015} derived a fifth-order Kadomtsev-Petviashvili (KP) equation in the long-wave approximation. It is noted that the KP equation was proposed for uni-directional surface waves with weak transverse variations. It cannot correctly describe the two-way propagation of surface waves, such as wave reflection/transmission across the media interface and head-on collisions of solitary waves. Additionally, it is not suitable to use an anisotropic model to study wave phenomena whose transverse variations are similar to those in the primary direction of propagation, such as the Kelvin wake and oblique interactions between lumps. Thus, though anisotropic uni-directional models have been successful in many aspects of free-surface problems, isotropic bi-directional models are still needed in some cases. 

Without the elastic cover, the isotropic Benney-Luke equation was derived originally by  \citet{Benney1964} to describe the bi-directional evolution of 3D small-amplitude pure gravity waves when the horizontal length scale is long compared with the fluid depth. \citet{Pego1999} rigorously proved the existence of finite-energy solitary waves for the Benney-Luke equation, including the effect of surface tension. \citet{Berger2000} established an explicit connection between the KP-I equation and the Benney-Luke equation considering strong surface tension, and studied the dynamics of lumps numerically. In many situations, the fluid depth is small compared with the typical wavelength of hydroelastic waves (e.g., see the experiment in Lake Saroma in Japan \cite{Takizawa1985}). Therefore, to simplify the primitive equations, it is reasonable to use the long-wave approximation and extend the Benney-Luke equation to the hydroelastic wave problem. 

The paper is structured as follows. In \S2, the detailed derivations of the Benney-Luke-type model and the associated envelope equation are presented. We then discuss the results of the envelope equation that served as a theoretical underpinning for hydroelastic solitary waves. In \S3, we numerically show different types of solitary waves in the Benney-Luke-type equation, together with their stability properties and collision dynamics. Besides, we compute the responses exerted by a locally confined load moving with a uniform speed, which features the shedding of lumps when the load speed is in the transcritical regime. A conclusion is given in \S4.

\section{Formulation}
\subsection{The Benney-Luke-type equation}
We consider a three-dimensional incompressible and inviscid fluid of density $\rho_w$ covered by a deformable ice sheet of density $\rho_i$ and the thickness $d$. The displacement of the ice cover is denoted by $z=\eta(x,y,t)$, where $x$ and $y$ are horizontal coordinates, and the $z$-axis points upwards with $z=0$ the undisturbed ice sheet. The fluid is bounded below by a rigid bed located at $z=-h$, where $h$ is constant. The motion of the fluid is assumed to be irrotational; therefore, we can introduce velocity potential, $\phi$, which satisfies the Laplace equation
\begin{equation}\label{Laplace}
\phi_{xx}+\phi_{yy}+\phi_{zz}=0\,,\quad\textrm{for}\;-h<z<\eta\left(x,y,t\right)\,.
\end{equation}
At the free surface $z=\eta\left(x,y,t\right)$, the kinematic and dynamic boundary conditions read 
\begin{equation}\label{kin1}
\eta_t = \phi_z-\left(\eta_x\phi_x+\eta_y\phi_y\right)
\end{equation}
and
\begin{equation}\label{dyn1}
\phi_t+\frac{1}{2}\left(\phi_x^2+\phi_y^2+\phi_z^2\right)+g\eta+\frac{P}{\rho_w}=p\left(x,y,t\right)\,,
\end{equation}
where $g$ is the acceleration due to gravity, $P$ the restoring force due to elastic bending, and $p(x,y,t)$ the external pressure exerted on the ice sheet (if $p<0$, the pressure acts downwards). The pressure resulting from the linear elastic bending model takes the form
\begin{equation}
P=D\Delta^2\eta + \rho_id\eta_{tt}\,,
\end{equation}
where $D=\frac{Ed^3}{12\left(1-\nu^2\right)}$ is the flexural rigidity ($E$ is Young's modulus and $\nu$ is the Poisson ratio), and $\Delta^2= \partial_{xxxx}+\partial_{yyyy}+2\partial_{xxyy}$ is the bi-Laplacian operator acting on horizontal variables. Hence the dynamic condition can be rewritten as
\begin{equation}\label{dyn2}
\phi_t+\frac{1}{2}\left(\phi_x^2+\phi_y^2+\phi_z^2\right)+g\eta+\frac{D}{\rho_w}\Delta^2\eta+\frac{\rho_id}{\rho_w}\eta_{tt}=p\,.
\end{equation}
Finally, the impermeability boundary condition at the bottom, 
\begin{equation}\label{kinbot}
\phi_z=0\,,\quad\text{at $z=-h$}\,,
\end{equation}
completes the whole system.

\citet{Ablowitz2006} introduced an explicit non-local formulation of the classical equations of water waves in both two and three dimensions. Following their work, we will derive a Benney-Luke-type equation for hydroelastic waves in the shallow-water regime in the subsequent analyses.

We reformulate these equations in terms of the surface displacement and surface velocity potential. If we denote by $q(x,y,t)=\phi(x,y,\eta(x,y,t),t)$ the value of $\phi$ at the free surface, then the dynamic boundary condition can be recast to
\begin{equation}\label{dyn3}
q_t+\frac{1}{2}|\nabla{q}|^2+g\eta-\frac{\left(\eta_t+\nabla q\cdot\nabla\eta\right)^2}{2\left(1+|\nabla\eta|^2\right)}+\frac{D}{\rho_w}\Delta^2\eta+\frac{\rho_id}{\rho_w}\eta_{tt}=p\,,
\end{equation}
where $\nabla$ is the horizontal gradient operator. For arbitrary harmonic functions $\phi$ and $\psi$, it is straightforward to verify that 
\begin{equation}\label{harmonic}
\left(\phi_z\psi_x+\psi_z\phi_x\right)_x+\left(\phi_z\psi_y+\psi_z\phi_y\right)_y+\left(\phi_z\psi_z-\psi_x\phi_x-\psi_y\phi_y\right)_z=0\,.
\end{equation}
We choose $\psi=e^{\ti\left(k_1x+k_2y\right)+kz}$, where $k=\sqrt{k_1^2+k_2^2}$ is the magnitude of the wavenumber vector. Substituting this solution into Eq. \eqref{harmonic}, integrating over the space occupied by the fluid, and applying the divergence theorem, one obtains the global relation
\begin{equation}\label{global}
\begin{aligned}
0=&\,\int_{\partial D} e^{\ti(k_1x+k_2y)+kz}\left[(\ti k_1\phi_z+k\phi_x)n_1+(\ti k_2\phi_z+k\phi_y)n_2\right.\\
&\,\left.+(k\phi_z-\ti k_1\phi_x-\ti k_2\phi_y)n_3\right]\mathrm{d}S\,,
\end{aligned}
\end{equation}
where $(n_1,n_2,n_3)^\top$ is the outward unit normal vector of $\partial D$, the boundary of the fluid body, and $\mathrm{d}S$ is the surface element. Using the kinematic boundary condition at the free surface and the impermeability condition at the bottom, Eq. \eqref{global} can be recast to
\begin{equation}\label{Fourier1}
\begin{aligned}
\iint_{\mathbb{R}^2}e^{\ti(k_{1}x+k_{2}y)}\Big[e^{k(\eta+h)}(k\eta_t-\ti k_{1}q_{x}-\ti k_{2}q_{y})+(\ti k_1\phi_x+\ti k_2\phi_y)_{z=-h}\Big]\mathrm{d}x\mathrm{d}y=0\,.
\end{aligned}
\end{equation}
In the same vein, substituting $\psi=e^{\ti(k_1x+k_2y)-kz}$ into Eq.~\eqref{harmonic} yields
\begin{equation}\label{Fourier2}
\begin{aligned}
\iint_{\mathbb{R}^2}e^{\ti(k_{1}x+k_{2}y)}\Big[-e^{-k(\eta+h)}(k\eta_t+\ti k_{1}q_{x}+\ti k_{2}q_{y})+(\ti k_1\phi_x+\ti k_2\phi_y)_{z=-h}\Big]\mathrm{d}x\mathrm{d}y=0\,.
\end{aligned}
\end{equation}
Subtracting Eq.~\eqref{Fourier2} from Eq.~\eqref{Fourier1} gives
\begin{equation}\label{Fourier3}
\begin{aligned}
\iint_{\mathbb{R}^2}\left\{\ti\eta_t\cosh[k(\eta+h)]+(\textbf{k}\cdot\nabla{q})\frac{\sinh\left[k(\eta+h)\right]}{k}\right\}e^{\ti(k_1x+k_2y)} \mathrm{d}x\mathrm{d}y=0\,,
\end{aligned}
\end{equation}
where we denote by $\bold{k}=(k_1,k_2)^\top$ the wavenumber vector. 

It is convenient to non-dimensionalize the problem using the Boussinesq scaling to proceed with the derivation. In this respect, we introduce the typical wavelength $l$, the small amplitude $a$, and the shallow-water speed $c_0=\sqrt{gh}$, and rescale the variables as
\begin{equation*}
\begin{aligned}
&x=lx^*\,,\quad y=ly^*\,,\quad k_1=\frac{k_1^*}{l}\,,\quad k_2=\frac{k_2^*}{l}\,,\\
&t=\frac{lt^*}{c_0}\,,\quad q=\frac{gla}{c_0}q^*\,,\quad \eta=a\eta^*\,,\quad p=\frac{p^*ga^2}{h}\,.
\end{aligned}
\end{equation*}
Then the dimensionless dynamic boundary condition and global relation read
\begin{equation}\label{dyn4}
q_t+\frac{\epsilon}{2}|\nabla q|^2+\eta-\frac{\epsilon\mu^2}{2}\frac{(\eta_t+\epsilon\nabla q\cdot\nabla\eta)^2}{1+\epsilon^2\mu^2|\nabla\eta|^2}+\delta\Delta^2\eta+\gamma\eta_{tt}=\epsilon{p}
\end{equation}
and
\begin{equation}\label{Fourier4}
\begin{aligned}
0=\iint_{\mathbb{R}^2}e^{\ti(k_1x+k_2y)}\left\{\ti\eta_t \cosh\left[\mu k(1+\epsilon\eta)\right]+(\textbf{k}\cdot\nabla q)\frac{\sinh\left[\mu k(1+\epsilon\eta)\right]}{\mu k}\right\}\mathrm{d}x\mathrm{d}y\,,
\end{aligned}
\end{equation}
respectively, where asterisks have been dropped for the ease of notations, and the dimensionless parameters read
\begin{equation*}
\epsilon=\frac{a}{h}\,,\quad\mu=\frac{h}{l}\,,\quad\delta=\frac{Ed^3}{12\rho_wg(1-\nu^2)l^4}\,,\quad\gamma=\frac{{\rho_i}dh}{{\rho_w}l^2}\,.
\end{equation*}
Let $\epsilon,\delta,\gamma=O(\mu^2)$ and this assumption corresponds to $a=0.1\,\mathrm{m}$, $d=1\,\mathrm{m}$, $h=10\,\mathrm{m}$, $l=100\,\mathrm{m}$, and $E=10^9\,\mathrm{N/m}^2$. It is noted that in the real situation, $\gamma=\frac{\rho_i}{\rho_w}\frac{d}{h}\mu^2$ is usually much smaller than $\mu^2$ due to the smallness of $d/h$; however, it does not qualitatively change the model equation if we involve the inertia effect in the current problem.

We expand the dynamic condition \eqref{dyn4} and the global relation \eqref{Fourier4} with respect to small parameters and retain terms valid up to $O(\epsilon,\delta,\gamma,\mu^2)$ to achieve a balance between dispersion and nonlinearity. After some algebra, one obtains the following Boussinesq-type equations
\begin{equation}\label{cutdyn}
\eta=-q_t-\frac{\epsilon}{2}|\nabla q|^2-\delta \Delta^2\eta-\gamma\eta_{tt}+\epsilon{p}\,,
\end{equation}
\begin{equation}\label{cutFourier}
\left(1-\frac{\mu^2}{2}\Delta\right)\eta_t+\left(\Delta-\frac{\mu^2}{6}\Delta^2\right)q
	+\epsilon\left(\nabla q\cdot\nabla\eta\right)+\epsilon\eta\Delta q=0\,,
\end{equation}
where $\Delta=\partial_{xx}+\partial_{yy}$ is the horizontal Laplace operator, and Eq. \eqref{cutFourier} is obtained by taking the inverse Fourier transform. By combining Eqs. \eqref{cutdyn} and \eqref{cutFourier} to eliminate $\eta$ and $\eta_t$, one arrives at
\begin{equation}\label{BL1}
\left[1-\left(\frac{\mu^2}{2}\Delta+\delta\Delta^2\right)\right]q_{tt}-\left(\Delta-\frac{\mu^2}{6}\Delta^2\right)q-\gamma q_{tttt}+\epsilon\left(\partial_t |\nabla q|^2+q_t\Delta q\right)-\epsilon{p_t}=0\,.
\end{equation}
Upon noticing $q_{tt}\sim\Delta{q}$ to leading order, Eq. \eqref{BL1} can be reformulated to
\begin{equation}\label{BL2}
q_{tt}-\Delta{q}-\left(\frac{\mu^2}{3}+\gamma\right)\Delta^2q-\delta\Delta^3q+\epsilon\left(\partial_t|\nabla{q}|^2+q_t\Delta{q}\right)=\epsilon{p_t}\,,
\end{equation}
a Benney-Luke-type equation with a time-dependent external force. 

By letting $p=0$ in Eq. \eqref{BL2}, we can derive the fifth-order Kadomtsev-Petviashvili (KP) equation, a single evolution equation describing uni-directional waves with weak variations in the transverse direction. For this purpose, we seek a solution of the form $q(x,y,t)=f(\bar{x},\bar{y},\bar{\tau})$, where $\bar{\tau}=\frac12\mu^2t$, $\bar{x}=x-t$, and $\bar{y}=\mu y$. In this asymptotic limit, we can rewrite Eq. \eqref{BL2} as
\begin{eqnarray*}
f_{\bar{x}\bar\tau}+f_{\bar{y}\bar{y}}+\left(\frac{1}{3}+\frac{\gamma}{\mu^2}\right)\partial_{\bar{x}}^4f+\frac{\delta}{\mu^2}\partial_{\bar{x}}^6f+\frac{3\epsilon}{\mu^2}f_{\bar{x}}f_{{\bar{x}{\bar{x}}}}=0
\end{eqnarray*}
upon neglecting higher-order terms. Differentiating the above equation with respect to $\bar{x}$ and substituting $\eta=f_{\bar{x}}+O\left(\mu^2\right)$, we obtain the fifth-order KP equation
\begin{equation}\label{KP} 
\left[\eta_{\bar\tau}+\left(\frac{1}{3}+\frac{\gamma}{\mu^2}\right)\partial_{\bar{x}}^3\eta+\frac{\delta}{\mu^2}\partial_{\bar{x}}^5\eta+\frac{3\epsilon}{\mu^2}\eta\eta_{\bar{x}}\right]_{\bar{x}}+\eta_{\bar{y}\bar{y}}=0\,.
\end{equation}
In the context of hydroelastic waves, the fifth-order KP equation \eqref{KP} was first obtained by \citet{Haragus1998} via a regular asymptotic expansion and re-derived by \citet{Guyenne2015} based on the Taylor expansion of the Dirichlet-Neumann operator in the Hamiltonian framework. If the $y$-dependence is negligible, Eq. \eqref{KP} reduces to the fifth-order Korteweg-de Vries equation, which is known as a reduced model for hydroelastic waves in channels (see \cite{Xia2002} for more details).

Finally, we show that without the external forcing, Eq. \eqref{BL2} has a Hamiltonian structure with the Hamiltonian functional
\begin{equation}\label{BLHam}
\begin{aligned}
\mathcal{H}[q,\xi]=\frac{1}{2}\iint_{\mathbb{R}^2}\left[\left(\xi-\frac{\epsilon}{2}|\nabla{q}|^2\right)^2+|\nabla{q}|^2+\delta|\nabla\Delta{q}|^2-\left(\frac{\mu^2}{3}+\gamma\right)(\Delta{q})^2\right]\mathrm{d}x\mathrm{d}y\,.
\end{aligned}
\end{equation}
The canonical variables are $q$ and $\xi$, where $\xi$ is defined as
\begin{equation}\label{xi}
\xi=q_t+\frac{\epsilon}{2}|\nabla{q}|^2\,.
\end{equation}
Calculating the variational derivatives with respect to the canonical variables gives
\begin{eqnarray}\label{qt}
\frac{\delta\mathcal{H}}{\delta\xi}=\xi-\frac{\epsilon}{2}|\nabla{q}|^2=q_t
\end{eqnarray}
and
\begin{equation}\label{xit}
\begin{aligned}
\frac{\delta\mathcal{H}}{\delta{q}}=&\,\epsilon\nabla\cdot\left[\left(\xi-\frac{\epsilon}{2}|\nabla{q}|^2\right)\nabla{q}\right]-\Delta{q}-\left(\frac{\mu^2}{3}+\gamma\right)\Delta^2{q}-\delta\Delta^3{q}\\
=&\,\epsilon\left[q_t\Delta{q}+\frac{1}{2}\left(|\nabla{q}|^2\right)_{t}\right]-\Delta{q}-\left(\frac{\mu^2}{3}+\gamma\right)\Delta^2{q}-\delta\Delta^3{q}\\
=&\,-q_{tt}-\frac{\epsilon}{2}\left(|\nabla{q}|^2\right)_{t}=-\xi_t\,.
\end{aligned}
\end{equation}

\subsection{The Benney-Roskes-Davey-Stewartson system}
Traditionally, weakly nonlinear wavepacket dynamics are investigated using the cubic nonlinear Schr\"{o}dinger (NLS) equation, which describes the envelope evolution of a monochromatic carrier wave in deep water. In the case of finite depth, the wave envelope is strongly coupled with the induced mean flow, and their interactions are governed by the Benney-Roskes-Davey-Stewartson (BRDS) system. It is well known that the envelope equations play a key role in understanding the existence of solitary wave solutions and their stability properties for both gravity-capillary and flexural-gravity problems (see \cite{Milewski20132,Kim2005} for example). In the context of 3D hydroelastic waves in the shallow-water regime, we expect to gain some theoretical predictions from the BRDS system derived from Eq. \eqref{BL2} in the absence of external forcing. For this purpose, we expand the solution of Eq. \eqref{BL2} as
\begin{equation}\label{solution}
q=\alpha{q_0}+\alpha^2{q_1}+\alpha^3{q_2}+\cdots\,,
\end{equation}
where the assumption of small amplitude is taken through the expansion in powers of $\alpha$. Then the equation to leading order reads 
\begin{equation*}
q_{0tt}-\Delta{q_0}-\left(\frac{\mu^2}{3}+\gamma\right)\Delta^2{q_0}-\delta\Delta^3{q_0}=0\,,
\end{equation*}
whose real solution is given by
\begin{equation}\label{1th}
q_0=A(X,Y,T)e^{\ti\theta}+\text{c.c.}+M(X,Y,T)\,,
\end{equation}
where c.c. stands for complex conjugation, $T=\alpha{t}$, $X=\alpha{x}$ and $Y=\alpha{y}$ are slow variables, $\theta=kx-\omega{t}$ is the fast variable with the dispersion relation $\omega^2=k^2-\left(\frac{\mu^2}{3}+\gamma\right)k^4+\delta{k^6}$, and $M(X,Y,T)$ is a real slow-varying, mean term.

We substitute $\partial_t=-\omega\partial_\theta+\alpha\partial_{T}$, $\partial_{x}=k\partial_\theta+\alpha\partial_{X}$, and $\partial_{y}=\alpha\partial_{Y}$ into the Benney-Luke-type equation. At second order, one obtains
\begin{eqnarray}\label{2thcon}
\left[2\ti\omega{A_T}+2\ti kA_X-4\ti k^3\left(\frac{\mu^2}{3}
+\gamma\right)A_X+6\ti\delta{k^5}A_X\right]e^{\ti\theta}=0\,,
\end{eqnarray}
and solve $q_1$ for what remains, i.e., solve
\begin{equation*}
\mathcal{L}q_1 = -3\ti\epsilon{k^2}\omega{A^2}e^{2i\theta}+\text{c.c.}
\end{equation*}
with $\mathcal{L} = \left(\omega^2-k^2\right)\partial^2_{\theta}-\left(\frac{\mu^2}{3}+\gamma\right)k^4\partial^4_{\theta}-\delta{k^6\partial^6_{\theta}}$. The former equation implies that $A(X,Y,T)=A(X-c_\text{g}T,Y)$, where $c_\text{g}$ is called the group velocity defined as 
\begin{equation*}
c_\text{g}=\frac{\rm{d}\omega}{\rm{d}k}=\frac{1}{\omega}\left[k-2\left(\frac{\mu^2}{3}+\gamma\right)k^3+3\delta k^5\right]\,.
\end{equation*}
The latter equation can be solved by letting $q_1=Be^{2\ti\theta}+\text{c.c.}$, which gives
\begin{equation*}
B = \frac{-3\ti\epsilon\omega{k^2A^2}}{-4\omega^2+4k^2-16\left(\frac{\mu^2}{3}+\gamma\right)k^4+64\delta{k^6}}\,.
\end{equation*}
To obtain the BRDS system, we transform to a moving reference frame $X-c_{\text{g}}T$ and denote by $\tau=\alpha{T}$ a much slower time scale. Then to the third order, we can get
\begin{equation}\label{BRDS1}
\left(1-c_{\text{g}}^2\right)M_{XX}+M_{YY}=\chi(|A|^2)_X\,,
\end{equation}
\begin{equation}\label{BRDS2}
\ti A_{\tau}+\frac{\omega^{''}}{2}A_{XX}+\frac{c_{\text{g}}}{2k}A_{YY}-\epsilon{kAM_X}+\Xi|A|^2A=0\,,
\end{equation}
where
\begin{equation*}
\chi=-\epsilon(k^2c_{\text{g}}+2k\omega)\,,\qquad\Xi=\frac{3\epsilon^2\omega}{-4\left(\frac{\mu^2}{3}+\gamma\right)+20\delta{k^2}}\,.
\end{equation*}  
The BRDS system \eqref{BRDS1}--\eqref{BRDS2} is valid for arbitrary $k>0$; however, in the current context, we focus on the case $k=k_{\text{min}}$, where $k_{\min}$ is the wavenumber corresponding to the phase speed minimum. Since $c_{\text{g}}^2<1$, $\omega^{''}>0$, and $c_{\text{g}}/2k>0$, the system is of `elliptic-elliptic' type as the coefficients of $M_{XX}$, $M_{YY}$, $A_{XX}$, and $A_{YY}$ are of the same sign. Following \cite{Cipolatti1992}, the system's behavior then depends on the sign of the parameter $\Theta$ defined as 
\begin{equation*}
\Theta = \Xi-\frac{\epsilon{k}\chi}{1-c_{\text{g}}^2}\,.
\end{equation*}
It was proved by \citet{Cipolatti1992} that $\Theta>0$ is a sufficient condition for the BRDS system to have a localized ground state solution, and the system is therefore called `focussing'. For the case that transverse modulations are absent (i.e., $\partial_{YY}=0$), the BRDS system \eqref{BRDS1}--\eqref{BRDS2} degenerates to the one-dimensional NLS equation $\ti A_\tau+\frac{\omega''}{2}A_{XX}+\Theta|A|^2A=0$. It is well known that the simplest solution to the one-dimensional focussing NLS equation takes an exact formula: $A=\widetilde{a}\,\text{sech}(\widetilde{b}X)e^{\ti\Omega\tau}$ with $\widetilde{a}=\pm\sqrt{2\Omega/\Theta}$ and $\widetilde{b}=\sqrt{2\Omega/\omega''}$.

In infinite depth, the effect of the mean flow vanishes, and the evolution is governed by the two-dimensional NLS equation 
\begin{equation}\label{NLS2D}
\ti A_\tau+A_{XX}+A_{YY}+|A|^2A=0\,,
\end{equation}
where coefficients have all been normalized to unity. It was shown by Alfimov \textit{et al.} \cite{Alfimov1990} that for the nonlinear eigenvalue problem of Eq. \eqref{NLS2D}, 
\begin{equation*}
\left\{
\begin{aligned}
&\rho_{XX}+\rho_{YY}-\rho+\rho^3=0\,,\\
&\rho(X,Y+l)=\rho(X,Y)\,,\\
&\lim_{X\to\pm\infty}\rho(X,Y)=0\,,
\end{aligned}
\right.
\end{equation*}
there exists a one-parameter family of solutions $\rho_l(X,Y)$ for $l>2\pi/\sqrt{3}$, where $l$ represents the basic period in the $Y$-direction. In contrast to plane solitons, this new type of solution features a nontrivial variation in the transverse direction. As the basic period $l\rightarrow{2\pi/\sqrt{3}}+0$, the solution $\rho_l$ degenerates to the plane soliton solution, and hence the emergence of transversally periodic solitons is called a dimension-breaking bifurcation. Following the new branch by increasing the bifurcation parameter $l$, a fully localized solution with circular symmetry can be obtained as $l\to\infty$; that is to say, these new solutions are a bridge between plane solitons and lumps. Inspired by the results of the asymptotic model, \citet{Milewski2014} found transversally periodic solitary waves in the full gravity-capillary wave problem corresponding to its NLS counterpart. Additionally, these authors predicted that the threshold wavelength of transverse instability of a plane solitary wave coincides with the dimension-breaking bifurcation point.

Motivated by their work, we numerically explore various types of localized steady solutions to the BRDS system through the dimension-breaking bifurcation, which provide a solid underpinning for the existence of solitary waves in the Benney-Luke-type equation. Without loss of generality, we set $\mu=0.3$, $\epsilon=\mu^2$, $\gamma=\delta=\mu^3$, and as a result, the BRDS system is of focussing type since the coefficient $\Theta>0$. To find localized time harmonic solutions to the BRDS system, we assume $A=e^{\ti\Omega\tau}\rho(X,Y)$, and then the steady BRDS system in the non-local form can be written as
\begin{equation}\label{eigeneq}
\begin{aligned}
-\Omega\rho+\frac{\omega^{''}}{2}\rho_{XX}+\frac{c_{\text{g}}}{2k}\rho_{YY}-\vartheta\epsilon{k}\chi\widetilde\Delta^{-1}\left[\rho^2\right]_{XX}\rho+\Xi\rho^3=0\,,
\end{aligned}
\end{equation}
where $\widetilde\Delta=\left(1-c_{\text{g}}^2\right)\partial_{XX}+\partial_{YY}$ and $\vartheta=1$. The parameter $\vartheta$ varying from 0 to 1 is introduced to assist with computing steady solutions to Eq. \eqref{eigeneq}. The numerical process can be divided into three steps in general. First of all, by letting $\vartheta=0$, the problem can be reduced to the steady focussing cubic NLS equation. This equation, after normalization, has countably many localized radial solutions (see \cite{Wang2012} for example), but we focus only on computing the ground state solution. Secondly, the obtained ground state solution is taken as the initial guess, and the non-local term in Eq. \eqref{eigeneq} is included by using a numerical continuation in $\vartheta$ up to $\vartheta=1$, resulting in a lump solution to the BRDS system. The lump is computed using the standard pseudo-spectral method in a horizontally double periodic domain with small wavenumbers $k_1$ and $k_2$ in the $X$- and $Y$-direction, respectively. Finally, the dimension-breaking bifurcation curve can be obtained by using $k_2$ as the continuation parameter, and the interested readers are referred to \cite{Milewski20132} for more details on the numerical scheme. 

\begin{figure*}[htbp]
	\centering
	\subfigure[]{		
		\includegraphics[width=0.45\linewidth]{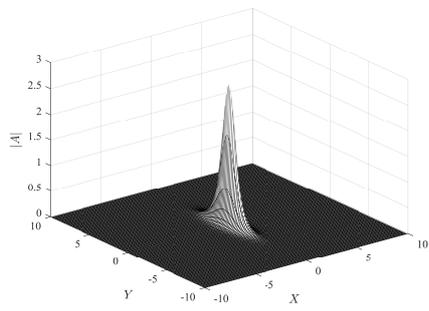}
		\label{fig:1a}	
	}
	\subfigure[]{
		\includegraphics[width=0.45\linewidth]{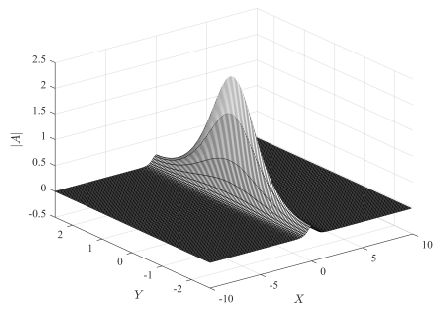}
		\label{fig:1b}	
	}
    \subfigure[]{
	    \includegraphics[width=0.42\linewidth]{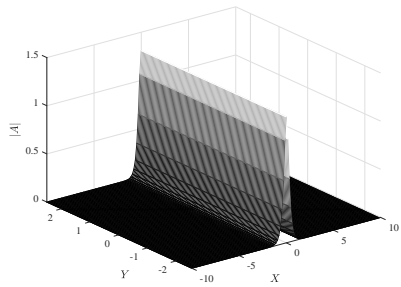}
	   \label{fig:1c}	
    }
	\caption{Typical solitons in the BRDS system: (a) lump; (b) transversally periodic soliton; (c) plane soliton.}
	\label{fig:1}
\end{figure*}

By gradually increasing $k_2$ (equivalently, decreasing the basic period $\bar{l}$ in the $Y$-direction), a one-parameter family of solutions $\rho_{\bar{l}}(X,Y)$ for $\bar{l}>2\pi/1.205$ is obtained. As the basic period $\bar{l}\to{2\pi/1.205+0}$, $\rho_{\bar{l}}$ degenerates to a plane soliton, while $\bar{l}\to\infty$, a lump solution emerges. The intermediate states which bridge the two limiting cases correspond to transversally periodic solitons. Fig. \ref{fig:1} shows the typical wave profiles on this new branch: \ref{fig:1a} and \ref{fig:1c} are the two limiting cases, while \ref{fig:1b} is the intermediate case. It is noted that the bifurcation wavenumber $k_2\approx1.205$ is closely related to the transverse instability of the plane soliton. 

\section{Results}
\subsection{Solitary waves}
\begin{figure*}[htbp]
\centering
\subfigure[]{		
\includegraphics[width=0.45\linewidth]{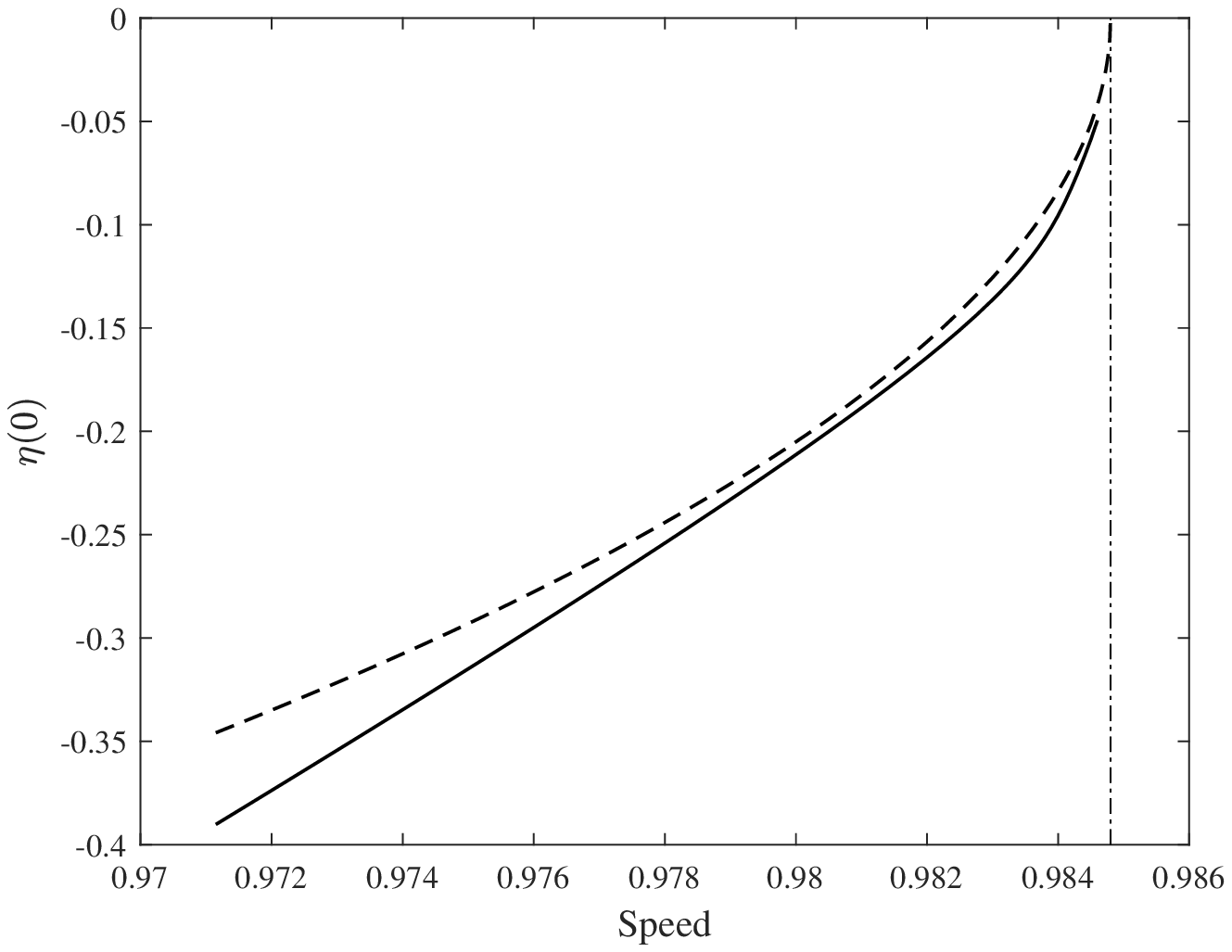}
\label{fig:2a}	
}
\subfigure[]{
	\includegraphics[width=0.45\linewidth]{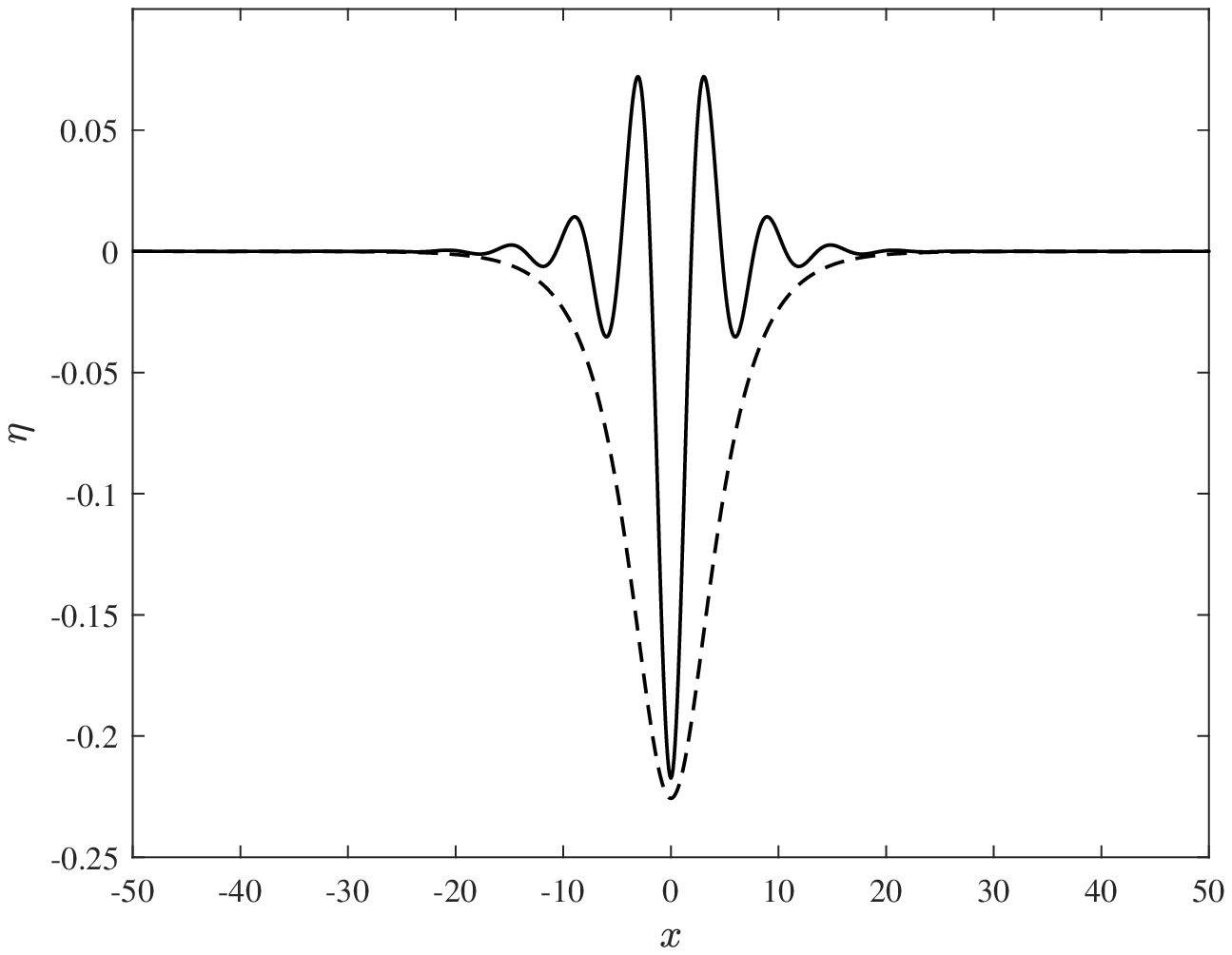}
	\label{fig:2b}	
}
\subfigure[]{		
	\includegraphics[width=0.45\linewidth]{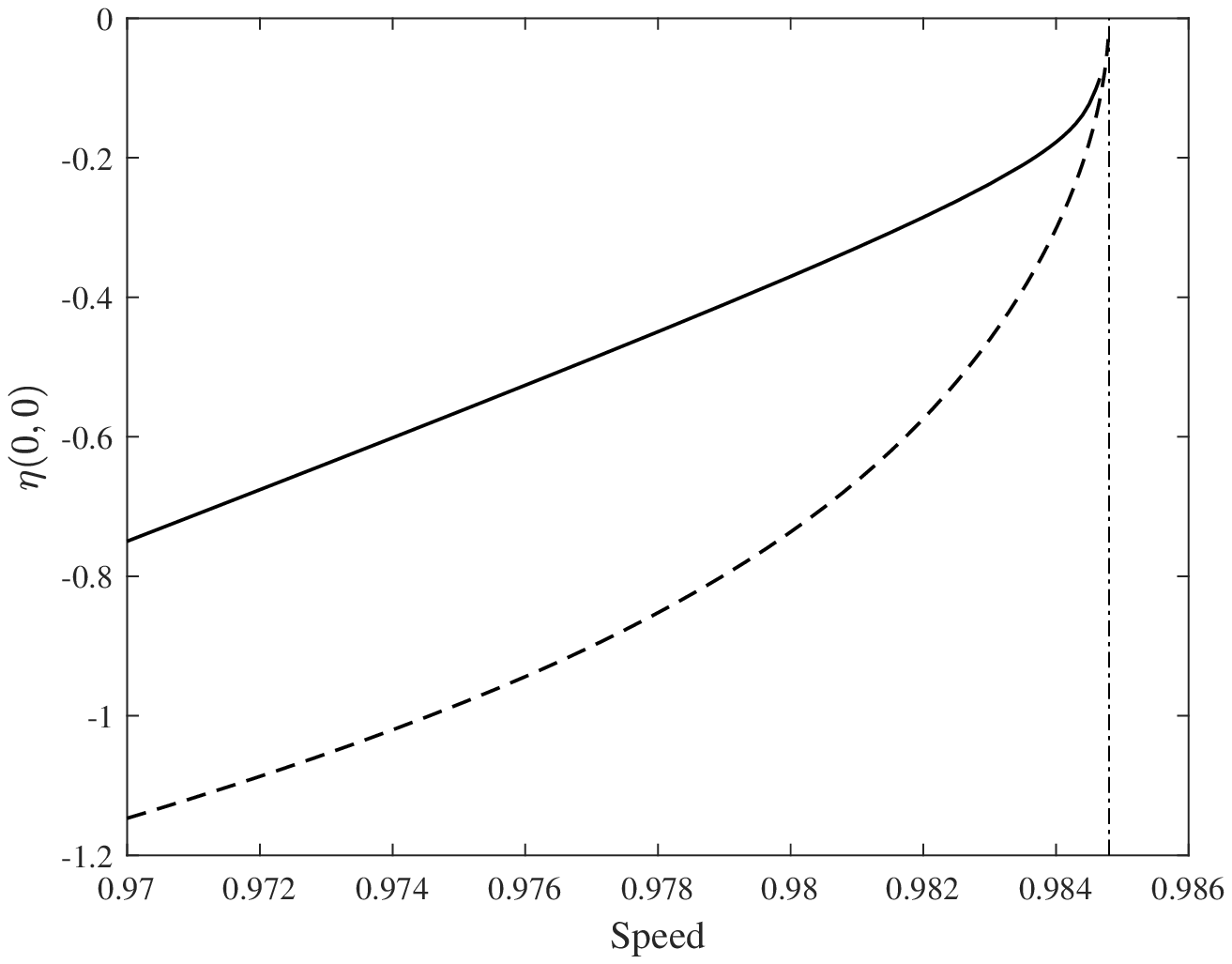}
	\label{fig:2c}	
}
\subfigure[]{
	\includegraphics[width=0.45\linewidth]{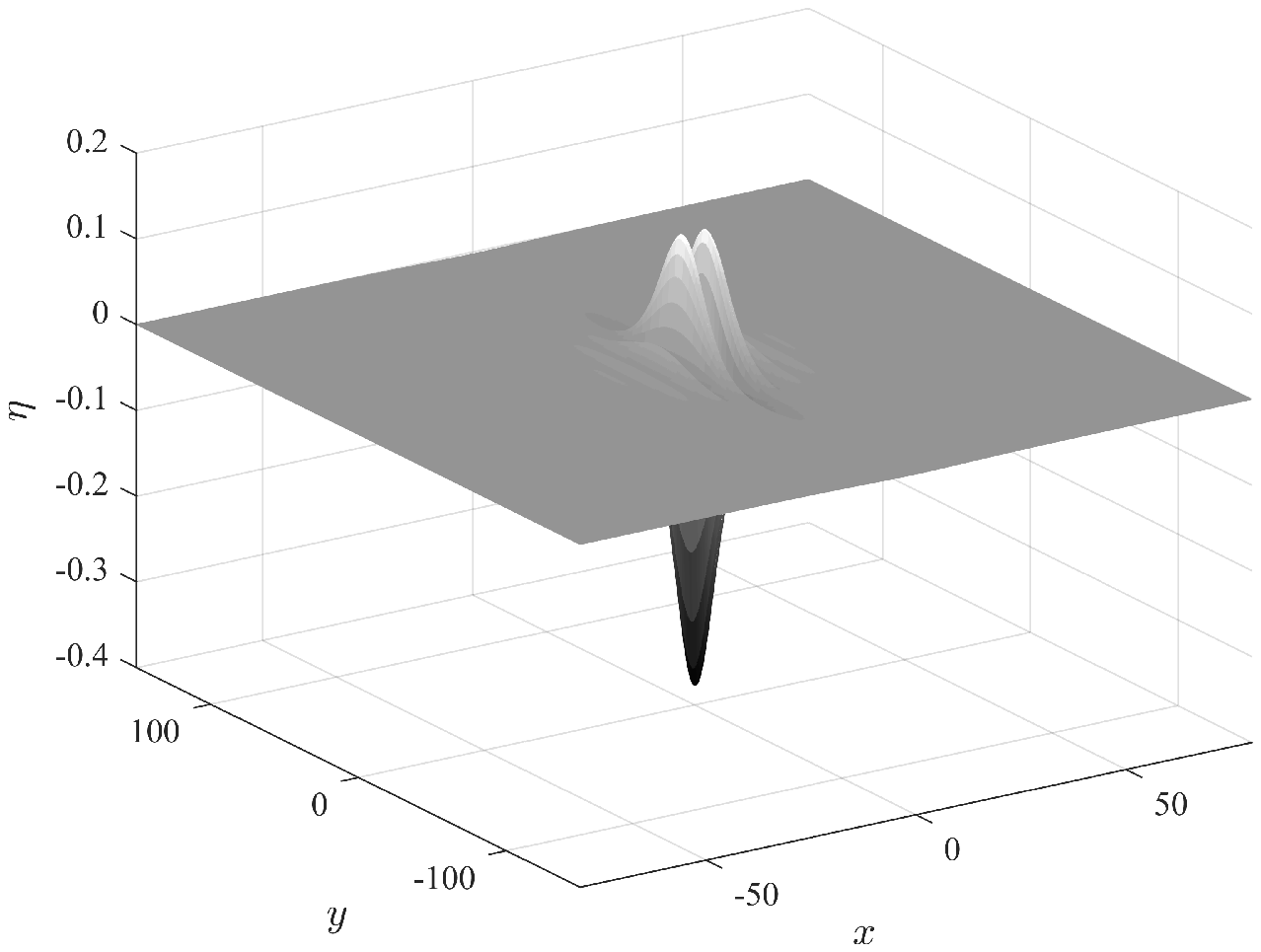}
	\label{fig:2d}	
}
\subfigure[]{		
	\includegraphics[width=0.45\linewidth]{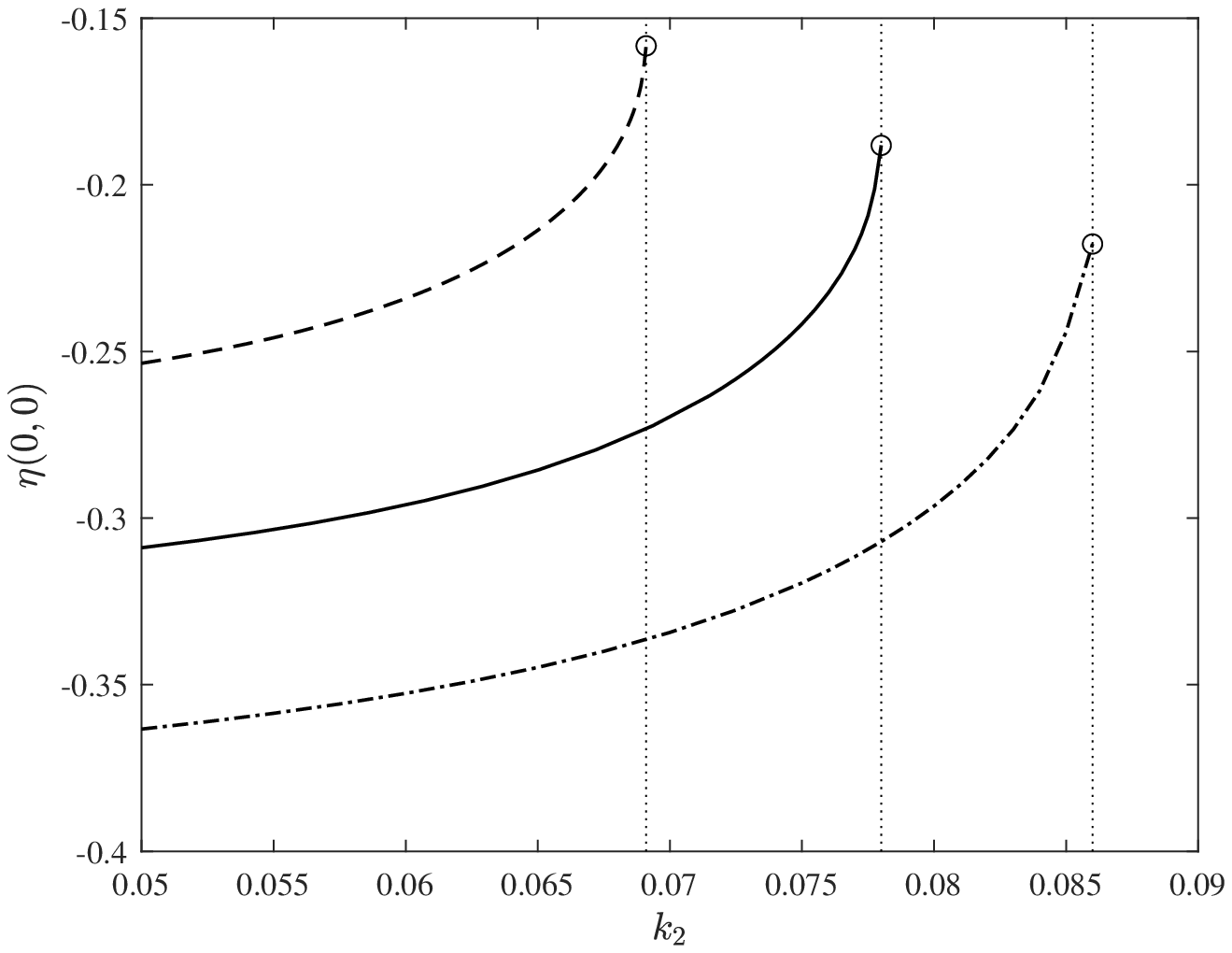}
	\label{fig:2e}	
}
\subfigure[]{
	\includegraphics[width=0.45\linewidth]{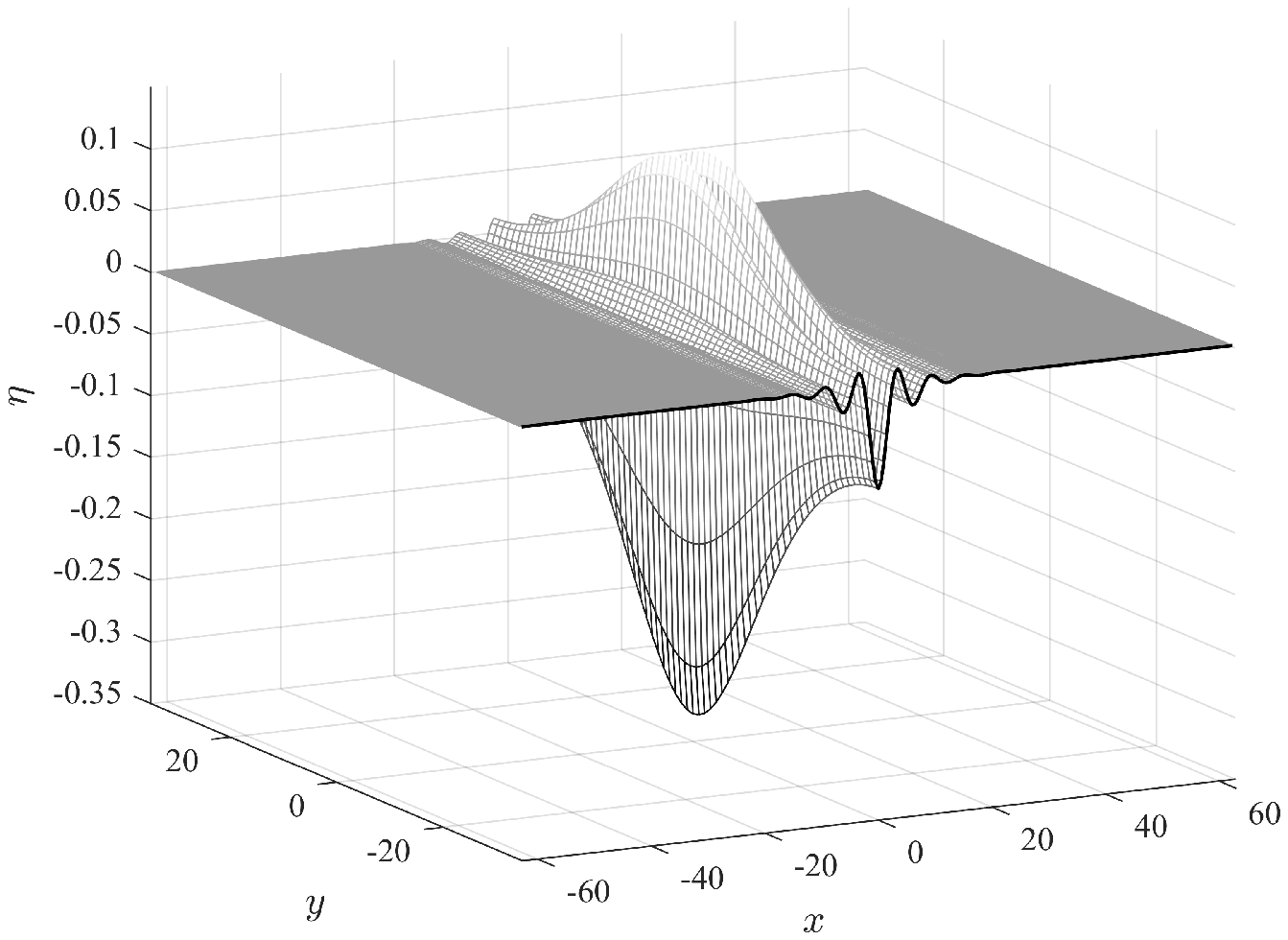}
	\label{fig:2f}	
}
\caption{(a) and (c): speed-amplitude bifurcation curves for plane solitary waves and lumps, respectively. Solid lines are associated with the Benney-Luke-type model and dashed lines are the BRDS predictions. The bifurcation speed is highlighted with vertical dash-dotted lines. (b) A typical plane solitary wave for $c=0.9798$ (solid line) is compared with the BRDS prediction (dashed line). (d) A typical lump profile for $c=0.9798$. (e) The dimension-breaking bifurcation curves for transversally periodic solitary waves with fixed speed: $c=0.9823$ (dashed line), $c=0.9811$ (solid line), and $c=0.9798$ (dash-dotted line). The dotted lines represent the bifurcation wavenumbers. (f) A transversally periodic solitary wave for $c=0.9798$ and $k_2=0.0725$. }
\label{fig:2}
\end{figure*}

As shown in Fig. \ref{fig:1}, three types of solitons (plane soliton, transversally periodic soliton, and lump) have been found in the BRDS system. Therefore we can expect their counterparts in the Benney-Luke-type equation (Eq. \eqref{BL2} without external forcing). To find solitary waves and the dimension-breaking bifurcation diagram, we start with computing lumps in the Benney-Luke-type equation. The numerical scheme is an extension of Petviashvili's method \cite{Petviashvili1976}, whose essence is to perform iterations in the Fourier space supplemented by a normalization factor. For this purpose, we first assume a lump is moving in an arbitrary horizontal direction, namely
\begin{equation*}
q(x,y,t)=q(x-c_{1}t,y-c_{2}t)\,,
\end{equation*}
where $c_1$ and $c_2$ are constants. Substituting this ansatz into Eq. \eqref{BL2} and performing the Fourier transform, one obtains
\begin{equation}
\hat{q}=\epsilon\frac{\ti(c_{1}k_{1}+c_{2}k_{2})\widehat{|\nabla{q}|^2}+c_{1}\widehat{q_{x}\Delta{q}}+c_{2}\widehat{q_{y}\Delta{q}}}{D}=\mathcal{P}\left[\hat{q}\right]\,,
\end{equation}
where
\begin{equation}
\begin{aligned}
D =k_1^2+k_2^2-\left(c_1k_1+c_2k_2\right)^2-\left(\frac{\mu^2}{3}+\gamma\right)\left(k_1^2+k_2^2\right)^2+\delta\left(k_1^2+k_2^2\right)^3\,,
\end{aligned}
\end{equation}
and hats denote the Fourier transform in $x$ and $y$. Following Ablowitz \textit{et al.} \cite{Ablowitz2006}, we introduce a multiplier in every iteration to prevent unlimited growth or reduction in amplitude, and hence the numerical scheme can be proposed as
\begin{equation}
\hat{q}_{n+1}=\alpha_n\mathcal{P}[\hat{q}_n]\,,\quad\text{with }\alpha_n=\dfrac{\displaystyle{\iint}{|\hat{q}_n|^2 \mathrm{d}k_1\mathrm{d}k_2}}{\displaystyle{\iint}{{\hat{q}_n}^*}\mathcal{P}[\hat{q}_n]\mathrm{d}k_1\mathrm{d}k_2}\,.
\end{equation}
In most numerical experiments, we compute solitary waves propagating in the positive $x$-direction, that is, $c_2=0$ and $c_1 = c$. In Fig. \ref{fig:2}, we present bifurcation curves and typical wave profiles of solitary waves in the Benney-Luke-type equation for $\mu=0.3$, $\epsilon=\mu^2$, and $\gamma=\delta=\mu^3$. The bifurcation diagrams \ref{fig:2a} and \ref{fig:2c} show the relation between the translating speed $c$ and the center of the free-surface displacement for plane solitary waves (locally confined in the direction of wave propagation without variations in the transverse direction) and lumps (localized in all horizontal directions), respectively. It is shown that both plane solitary waves and lumps bifurcate from $c\approx0.9848$, and only depression waves, which feature a negative free-surface elevation at their center, can be found for positive $c$. The dashed lines in \ref{fig:2a} and \ref{fig:2c} represent the bifurcation curves that take the leading-order approximation of amplitude $\eta\sim-{q_t}$ in the BRDS system, indicating that the theoretical predictions are valid only for a very narrow range of speed in the vicinity of the bifurcation point. The dimension-breaking bifurcation curves are plotted in Fig. \ref{fig:2e}, demonstrating the relationship between the basic wavenumber in the transverse direction and $\eta(0,0)$ for fixed $c=0.9798$, $0.9811$, $0.9823$ (from right to left). Typical examples of three types of solitary waves are plotted on the right column: plane solitary wave (\ref{fig:2b}), lump (\ref{fig:2d}), and transversally periodic solitary wave, which are localized in the propagation direction and periodic in the transverse direction (\ref{fig:2f}).

In the gravity-capillary (GC) wave problem, the KP-I equation proposed for unidirectional waves with weak transverse variation has rational solutions, which can help to understand the algebraic decay of lumps in the Benney-Luke equation with strong surface tension. However, regarding hydroelastic waves, no analytic travelling-wave solutions are known to the fifth-order KP equation. Alternatively, we manage to understand the far-field behavior of lumps of the Benney-Luke-type equation with the aid of the associated BRDS system. Using the envelope equations, \citet{Kim2005} showed that GC lumps feature algebraically decaying tails at infinity regardless of water depth owing to the algebraic decay of the induced mean flow. Following their argument, we first take the Fourier transform of the mean flow equation \eqref{BRDS1}, which yields
\begin{eqnarray*}
\frac{\partial{M}}{\partial{X}}=\chi\iint_{\mathbb{R}^2}\widehat{|A|^2}\,\frac{k_1^2\,e^{\ti(k_{1}X+k_{2}Y)}}{(1-c_{\text{g}}^2)k_1^2+k_2^2}\,\mathrm{d}k_1\mathrm{d}k_2\,.
\end{eqnarray*}
where $\widehat{|A|^2}$ is the Fourier transform of $|A|^2$ in $X$ and $Y$. Therefore, as $X^2+Y^2\to\infty$,
\begin{eqnarray*}
\frac{\partial{M}}{\partial{X}}\sim-\frac{I_0\chi}{2\pi\sqrt{1-c_{\text{g}}^2}}\frac{\partial}{\partial{Y}}\left[\frac{Y}{X^2+(1-c_{\text{g}}^2)Y^2}\right]\,,
\end{eqnarray*}
where
\begin{equation*}
I_0=\iint_{\mathbb{R}^2}|A|^2\mathrm{d}X\mathrm{d}Y\,.
\end{equation*}
This indicates that the induced mean flow decays algebraically at infinity, and the envelope $A$ of primary harmonic features exponential decaying tails according to Eq. \eqref{BRDS2}. As a result, the induced mean flow controls the behavior at the tails of a lump. The free surface elevation at the tails is obtained $\eta\sim\alpha^2cM_X$ and decays algebraically as well,
\begin{eqnarray*}
\eta\sim-\frac{\alpha^2cI_0\chi}{2\pi\sqrt{1-c_{\text{g}}^2}}\frac{\partial}{\partial{Y}}\left[\frac{Y}{X^2+(1-c_{\text{g}}^2)Y^2}\right]\,.
\end{eqnarray*}

\begin{figure*}[htbp]
	\centering
	\subfigure[]{		
		\includegraphics[width=0.45\linewidth]{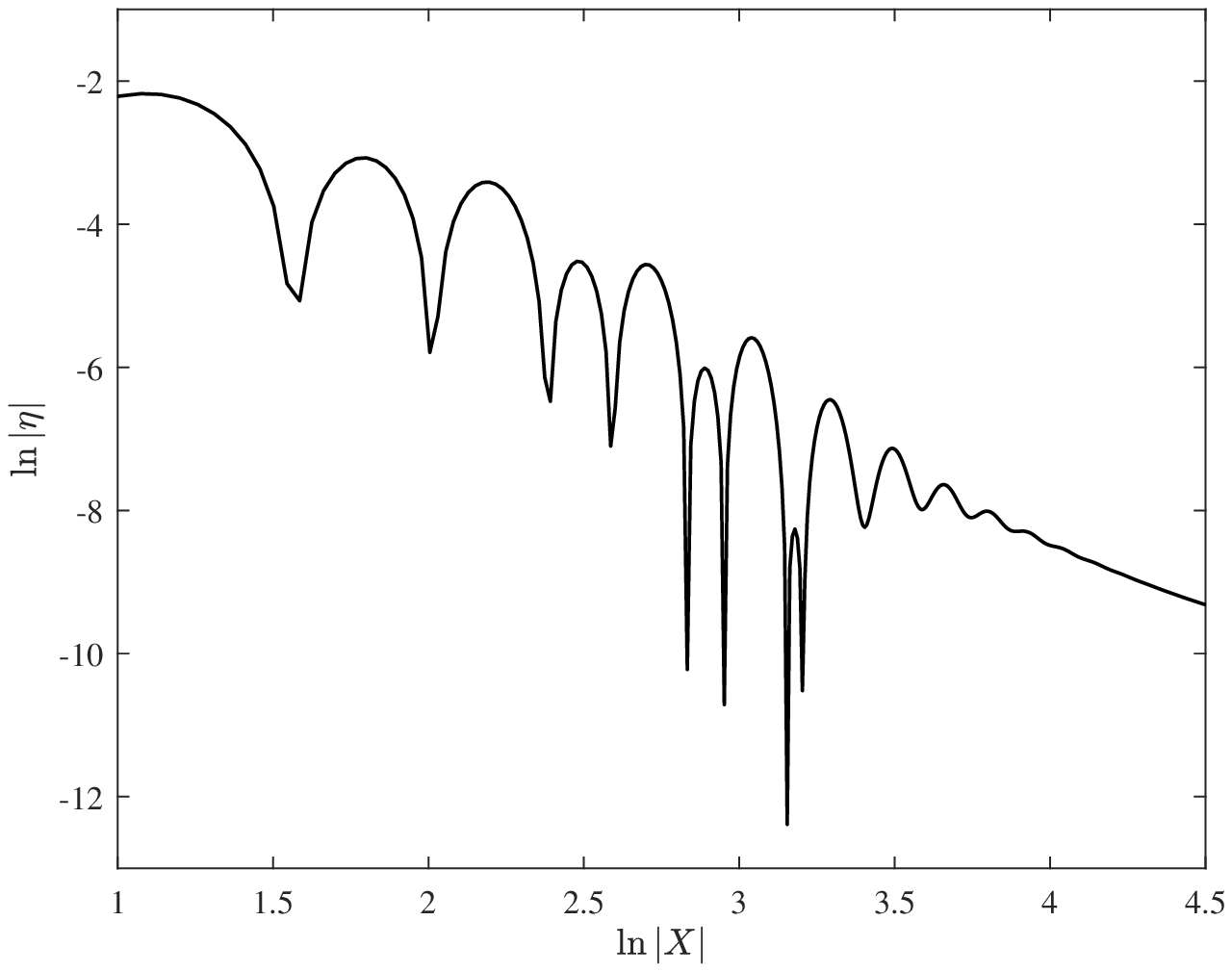}
		\label{fig:3a}	
	}
	\subfigure[]{
		\includegraphics[width=0.45\linewidth]{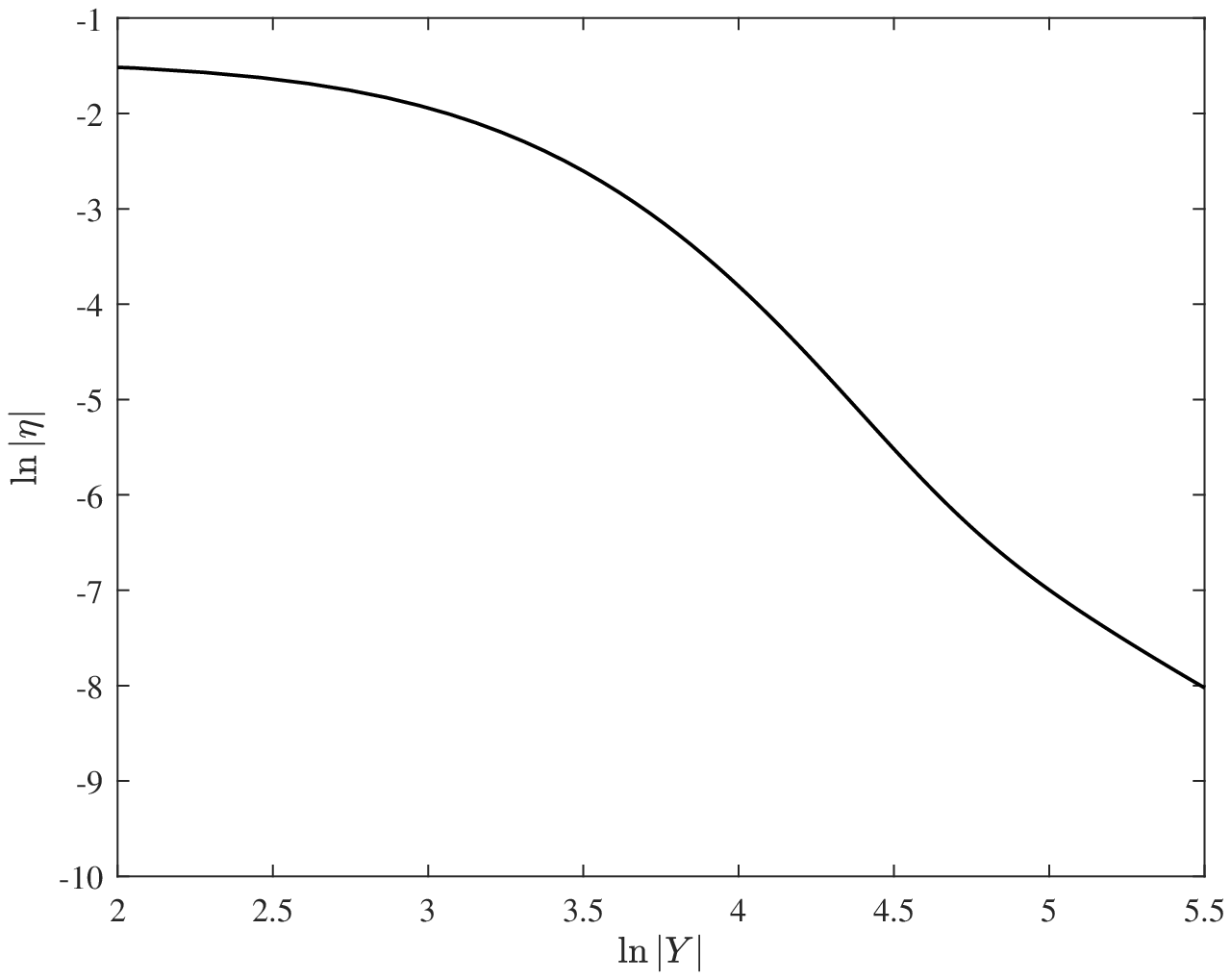}
		\label{fig:3b}	
	}
	\caption{Log-log plots of the free surface elevation $|\eta|$ as the function of $|X|$ and $|Y|$ for the depression lump with $c=0.983$.}
	\label{fig:3}
\end{figure*}

The numerical solutions to the Benney-Luke-type equation are presented in Fig. \ref{fig:3} to confirm the asymptotic results from the BRDS system. The plots of $\ln|\eta|$ against $\ln|X|$ and $\ln|Y|$ for the lump with the speed $c=0.983$ are shown. Note that, in both plots, the free-surface elevation approaches a straight line with slope $-2$ to a good approximation, which is consistent with the prediction that the tails ultimately decay algebraically like $X^{-2}$ and $Y^{-2}$ in the $x$- and $y$-direction, respectively.

\subsection{Transverse instability}
The stability of line solitary waves subject to transverse long-wavelength perturbations was explored extensively in gravity-capillary water-wave problems by different groups. \citet{Kim2007} provided a long wave stability analysis and concluded that the leading-order instability growth rate was associated with the change rate of the mechanical energy relative to the wave speed. However, their argument could not give the threshold wavelength of the transverse instability. On the other hand, the linear stability analysis of plane solitary waves of the 2D NLS equation indicated the critical wavenumber of transverse instability \cite{Rypdal2006}. Motivated by this result, \citet{Akers2009} and \citet{Wang2012} obtained the threshold wavelength of perturbations and performed direct numerical simulations to verify the theoretical prediction. The studies mentioned above are related to depression GC solitary waves, while the elevation ones were discussed by \citet{Wang2015}, and similar results were obtained. In the subsequent analyses, we first perform an asymptotic analysis of the instability of plane solitary waves subject to transverse long-wavelength perturbations and then specify the critical wavenumber of triggering instability.

\begin{figure}[htbp]
	\includegraphics[width=\linewidth]{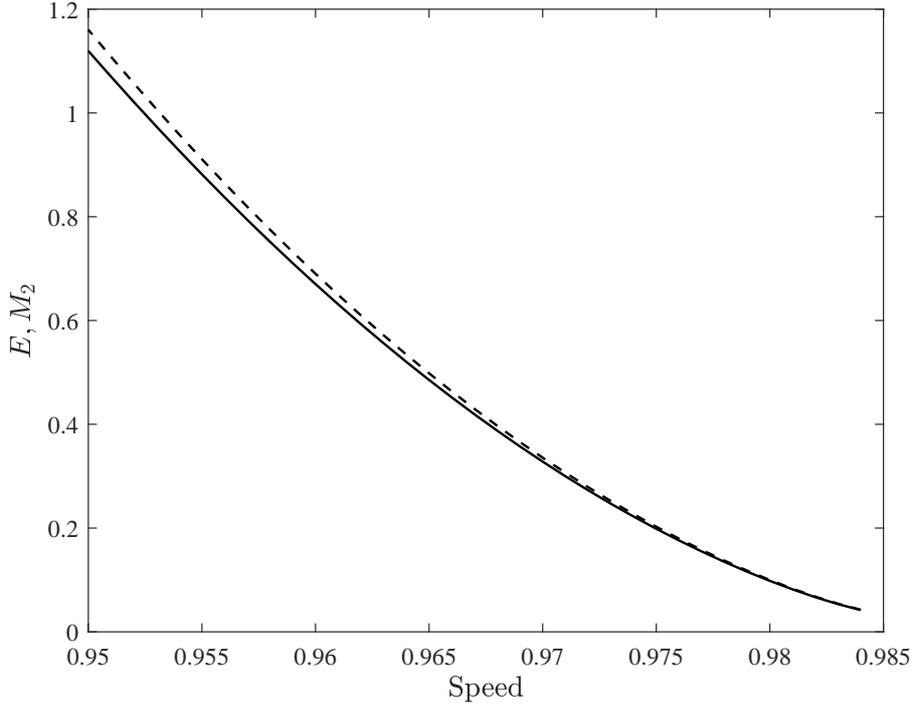}\
	\caption{$E$ (dashed line) and $M_2$ (solid line) versus the solitary-wave speed.}
	\label{fig:4} 
\end{figure}

We assume that $U({\sigma};c)$ is a plane solitary wave solution characterized by the translating speed $c$, where $\sigma=x-ct$. Substituting this solution into the unforced Benney-Luke-type equation yields
\begin{equation}\label{21}
(c^2-1)U_{\sigma\sigma}-\left(\frac{\mu^2}{3}+\gamma\right)\partial_{\sigma}^4U-\delta\partial_{\sigma}^6U-3c\epsilon U_\sigma U_{\sigma\sigma}=0\,.
\end{equation}
Taking the derivative of Eq. \eqref{21} with respect to $c$ yields
\begin{equation}\label{22}
-2cU_{\sigma\sigma}+3\epsilon{U_{\sigma}U_{\sigma\sigma}}=\left(c^2-1\right)U_{c\sigma\sigma}-\left(\frac{\mu^2}{3}+\gamma\right)\partial_\sigma^4U_c-\delta\partial_\sigma^6U_c-3c\epsilon\left({U_{\sigma}U_{c\sigma}}\right)_{\sigma}\,.
\end{equation}
We perturb the plane solitary wave solution in the transverse direction using a cosine longwave, namely $q=U(\sigma;c)+\tilde{q}(\sigma;c)e^{\ti\beta{y}+\lambda{t}}+\text{c.c.}$ with a small wavenumber $\beta$ in the direction transverse to wave propagation. Substituting the ansatz into the Benney-Luke-type equation and collecting the coefficients of $e^{\ti\beta{y}+\lambda{t}}$, one can obtain
\begin{equation}\label{23}
\begin{aligned}
0=&\,\left[\lambda^2+\beta^2-\left(\frac{\mu^2}{3}+\gamma\right)\beta^4+\delta\beta^6+\epsilon\beta^2cU_{\sigma}+\epsilon\lambda{U_{\sigma\sigma}}\right]\tilde{q}+\left(2\epsilon\lambda{U_\sigma}-3c\epsilon{U_{\sigma\sigma}}-2c\lambda\right){\tilde{q}}_\sigma\\
&\,+\left[c^2-1+2\beta^2\left(\frac{\mu^2}{3}+\gamma\right)-3\delta\beta^4-3c\epsilon{U_\sigma}\right]\tilde{q}_{\sigma\sigma}+\left[3\delta\beta^2-\left(\frac{\mu^2}{3}+\gamma\right)\right]\partial_{\sigma}^4\tilde{q}-\delta\partial_{\sigma}^6\tilde{q}\,.
\end{aligned}
\end{equation}
Furthermore, we assume
\begin{subequations}\label{24}
\begin{eqnarray}
\lambda&=&\beta\lambda^{(1)}+\beta^2\lambda^{(2)}+\beta^3\lambda^{(3)}+\cdots\,,\label{24a}\\
\tilde{q}&=&q^{(0)}+\beta{q^{(1)}}+\beta^2{q^{(2)}}+\cdots\,.\label{24b}
\end{eqnarray}
\end{subequations}
By substituting expansions \eqref{24a} and \eqref{24b} into Eq. \eqref{23} and equating like powers of $\beta$, we obtain, at O(1),
\begin{equation}\label{25}
0=\left(c^2-1\right)q^{(0)}_{\sigma\sigma}-\left(\frac{\mu^2}{3}+\gamma\right)\partial_\sigma^4 q^{(0)}-\delta\partial_\sigma^6q^{(0)}-3c\epsilon\left(U_{\sigma}q^{(0)}_{\sigma}\right)_{\sigma}\,.
\end{equation}
It is evident that $q^{(0)}=U_{\sigma}$ is a solution to Eq. \eqref{25}. To $O(\beta)$, $q^{(1)}$ satisfies the forced problem
\begin{equation}\label{26}
\lambda^{(1)}\left(2cU_{\sigma\sigma}-3\epsilon{U_{\sigma}U_{\sigma\sigma}}\right)=\left(c^2-1\right)q^{(1)}_{\sigma\sigma}-\left(\frac{\mu^2}{3}+\gamma\right)\partial_{\sigma}^4{q^{(1)}}-{\delta}\partial_{\sigma}^6{q^{(1)}}
-3c\epsilon\left(U_{\sigma}q^{(1)}_{\sigma}\right)_{\sigma}\,.
\end{equation}
Recalling \eqref{22}, the solution to Eq. \eqref{26} is $q^{(1)}=-\lambda^{(1)}U_c$. At the next order, the equation for $q^{(2)}$ reads
\begin{equation}\label{27}
\begin{aligned}
&\,\left[\left(c^2-1\right)\partial_{\sigma}^2-\left(\frac{\mu^2}{3}+\gamma\right)\partial_{\sigma}^4-{\delta}\partial_{\sigma}^6\right]{q^{(2)}}-3c\epsilon\left(U_{\sigma}q^{(2)}_\sigma\right)_{\sigma}\\
=&\,-\left({\lambda^{(1)}}^2+c\epsilon{U_\sigma}+1+\epsilon\lambda^{(2)}U_{\sigma\sigma}\right)q^{(0)}-\epsilon\lambda^{(1)}U_{\sigma\sigma}q^{(1)}-2\epsilon\lambda^{(2)}U_{\sigma}q^{(0)}_{\sigma}\\
&\,-2\epsilon\lambda^{(1)}U_{\sigma}q^{(1)}_{\sigma}+2c\lambda^{(1)}q^{(1)}_\sigma+2c\lambda^{(2)}q^{(0)}_{\sigma}-2\left(\frac{\mu^2}{3}+\gamma\right)q^{(0)}_{\sigma\sigma}-3{\delta}\partial_{\sigma}^4q^{(0)}\,.
\end{aligned}
\end{equation}
The adjoint operator of the left-hand side of Eq. \eqref{27} is
\begin{equation}\label{28}
	\mathcal{L^+}=(c^2-1)\partial^2_{\sigma}-\left(\frac{\mu^2}{3}+\gamma\right)\partial^4_{\sigma}
	-\delta\partial^6_{\sigma}-3c\epsilon(U_{\sigma}\partial_{\sigma})_{\sigma}\,,
\end{equation}
and it is obvious that $\mathcal{L^+}U_{\sigma}=0$. If we denote by $R$ the right-hand side of \eqref{27}, then the solvability condition for the inhomogeneous equation \eqref{27} is
\begin{equation*}
\int_{-\infty}^{\infty}{R}U_{\sigma}\mathrm{d}\sigma=0\,,
\end{equation*}
which gives
\begin{equation}\label{29}
\int_{-\infty}^{\infty}\left[U_{\sigma}^2+c\epsilon{U_{\sigma}^3}-2\left(\frac{\mu^2}{3}+\gamma\right)U_{\sigma\sigma}^2+3\delta{U_{\sigma\sigma\sigma}^2}\right]\mathrm{d}\sigma=\left({\lambda^{(1)}}\right)^2\int_{-\infty}^{\infty}\left(\frac{1}{2}\epsilon{U_{\sigma}^3-cU_{\sigma}^2}\right)_{c}\mathrm{d}\sigma\,.
\end{equation}
The integrals on the right and left sides are denoted by $M_1$ and $M_2$, respectively. We infer from Eq. \eqref{29} that the plane solitary wave $U$ is transversely unstable if
\begin{equation*}
M_1M_2>0\,.
\end{equation*}
By letting 
\begin{equation*}
E=\int_{-\infty}^{\infty}\left(cU_{\sigma}^2-\frac{1}{2}\epsilon{U_{\sigma}^3}\right)\mathrm{d}\sigma\,,
\end{equation*}
the values of $E$ and $M_2$ can be calculated numerically. As shown in Fig. \ref{fig:4}, $E$ and $M_2$ are both positive and decrease monotonically along with the translating speed $c$. Consequently, the instability condition is satisfied and plane solitary waves are proved to be transversally unstable under long-wavelength disturbances.

\begin{figure}[htbp]
	\includegraphics[width=\linewidth]{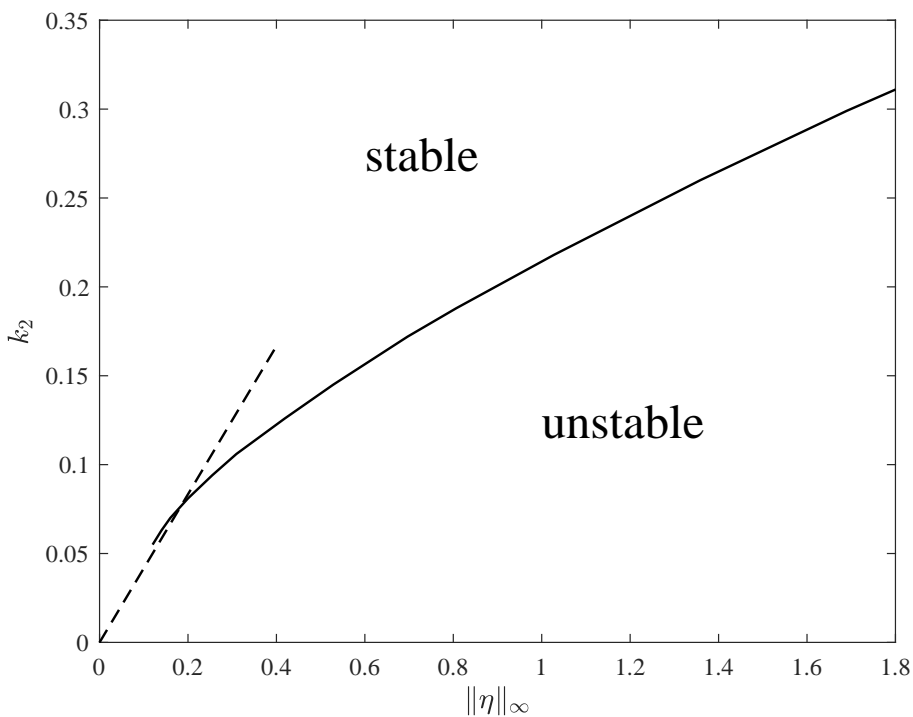}\
	\caption{Relation between the amplitude of the plane solitary wave and the critical perturbation wavenumber for triggering the transverse instability. Solid line: results obtained from numerical computations of the Benney-Luke-type model; dashed line: the BRDS prediction.}
	\label{fig:5} 
\end{figure}

It is mentioned in \S2 that the critical wavenumber for the onset of transverse instability of a plane solitary wave is coincident with the dimension-breaking bifurcation point. It is natural to establish the relation between the amplitude of plane solitary waves and the critical perturbation wavenumber $k_{\mathrm{c}}$ for the transverse instability in the Benney-Luke-type equation. It was shown in the last paragraph of \S2 that for the BRDS plane solitons, when the wavenumber in the $Y$-direction increases to 1.205, denoted as $K_\mathrm{c}$, the dimension-breaking phenomenon occurs. Using the relations $k_{\mathrm{c}}\approx\alpha{K_{\mathrm{c}}}$ and $||\eta||_{\infty}\approx2\alpha\omega|A|$ and eliminating $\alpha$ by using wave amplitude, we then obtain
\begin{eqnarray}\label{keta}
k_{\mathrm{c}}\approx0.4167||\eta||_{\infty}\,.
\end{eqnarray}
On the other hand, the critical wavenumber for the onset of transverse instability in the Benney-Luke-type equation can be obtained by direct numerical computations. We start with a lump solution and then gradually shrink the computational domain in the $y$-direction (by fixing the wave speed) until the variations in the transverse direction become trivial. Fig. \ref{fig:5} shows the relation between the amplitude of plane solitary waves and the dimension-breaking bifurcation point (solid line), indicating that the asymptotic prediction (dashed line) given by Eq. \eqref{keta} is valid only for small-amplitude waves. It then follows that the parameter space ($||\eta||_\infty$, $k_2$) shown in Fig. \ref{fig:5} can be divided into two regions: below and above the solid curve. For a given plane solitary wave (which can be determined by its amplitude), the transverse instability occurs when it is superimposed with a harmonic perturbation, orthogonal to the direction of wave propagation, with the wavenumber in the lower region. In contrast, it is stable when the perturbation wavenumber is in the upper region.

\begin{figure*}[htbp]
	\centering
	\subfigure[]{		
		\includegraphics[width=0.45\linewidth]{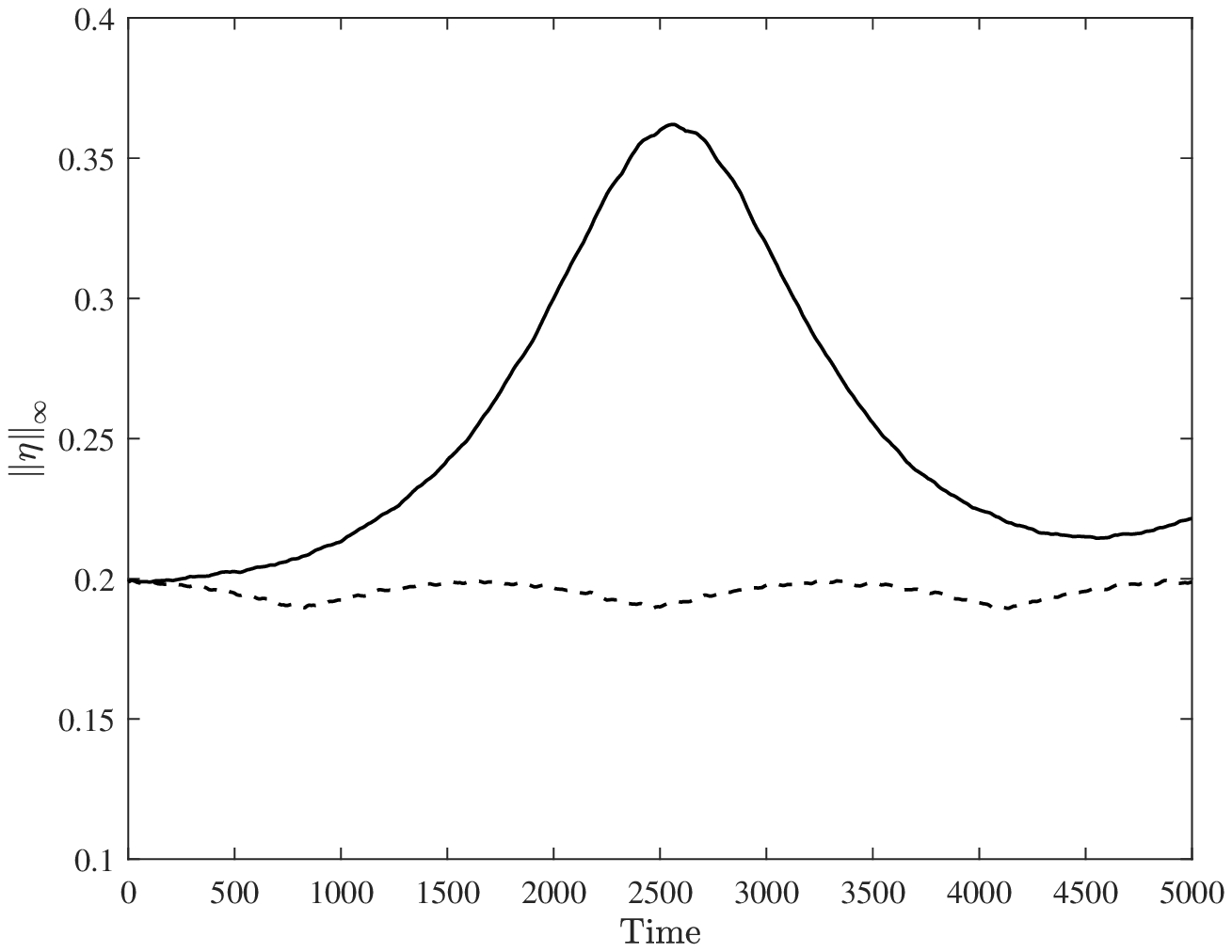}
		\label{fig:6a}	
	}
	\subfigure[]{
		\includegraphics[width=0.45\linewidth]{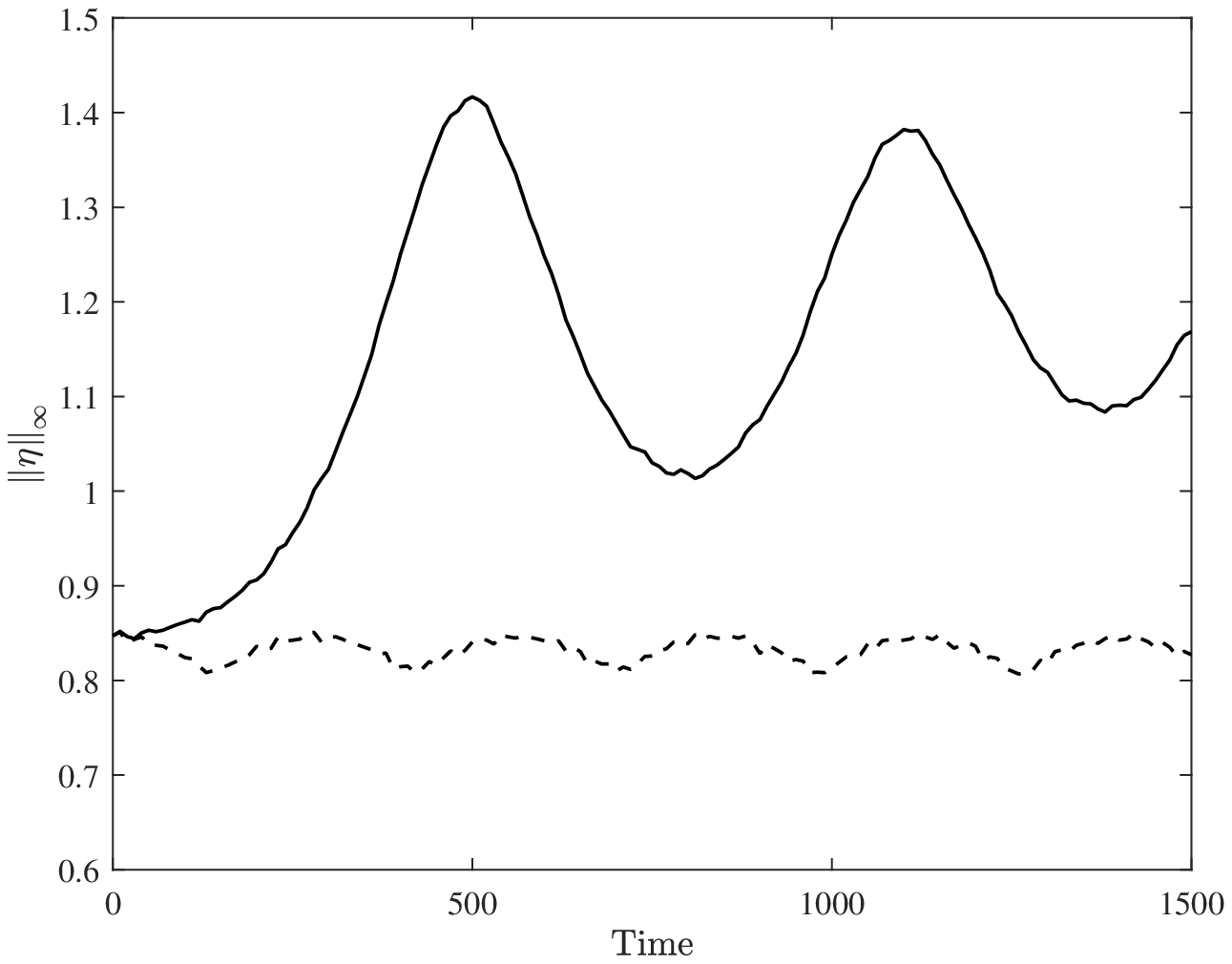}
		\label{fig:6b}	
	}
	\caption{Stability and instability of plane solitary waves subject to transverse perturbations. (a) A solitary wave of amplitude $0.1894$ is perturbed by a cosine function with $k_2=0.07$ (solid line) and $k_2=0.086$ (dashed line). (b) A solitary wave of amplitude $0.8069$ is perturbed by a cosine function with $k_2=0.168$ (solid line) and $k_2=0.208$ (dashed line).}
	\label{fig:6}
\end{figure*}

\begin{figure}[htbp]
	\centering
	\subfigure[]{
		\includegraphics[width=\linewidth]{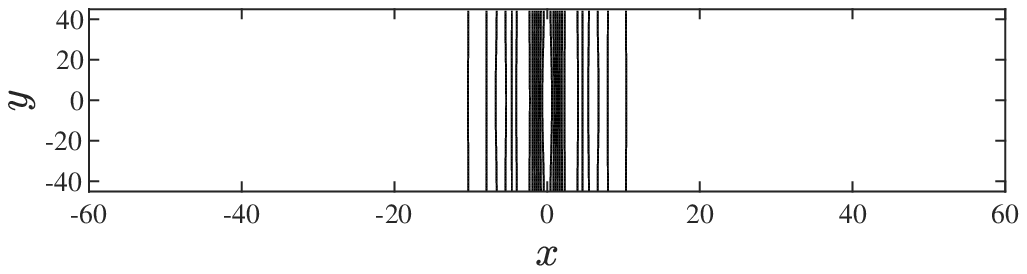}
		\label{fig:7a}}
	\subfigure[]{
		\includegraphics[width=\linewidth]{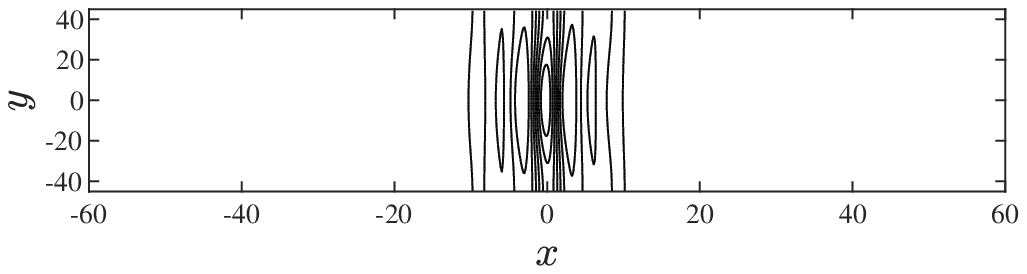}
		\label{fig:7b}}
	\subfigure[]{
		\includegraphics[width=\linewidth]{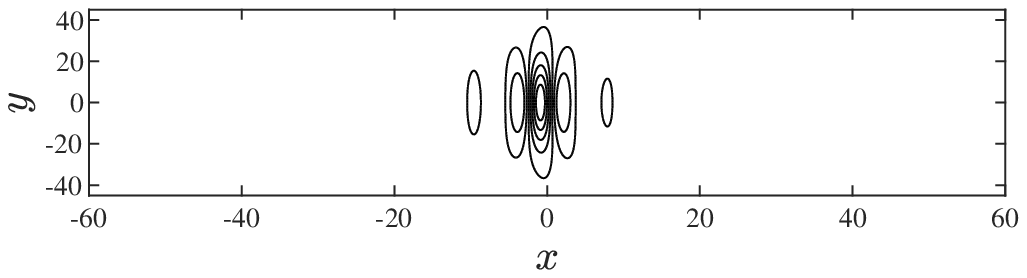}
		\label{fig:7c}}
	\caption{Time evolution of a plane solitary wave in the presence of transverse perturbation  (the small-amplitude case with $k_2=0.07$; see Fig. \ref{fig:6a}). The contours are plotted at (a) $t=0$, (b) $t=1500$, and (c) $t=2500$.}
	\label{fig:7}
\end{figure}

The stability properties obtained above are confirmed via numerical time evolutions of transversally perturbed plane solitary waves. We numerically integrate the Hamilton equations \eqref{qt}--\eqref{xit} using the pseudo-spectral algorithm proposed by \citet{Milewski1999}, who introduced symmetric factorization of the linear operator and applied the integrating factor method in the spectral space (the interested readers are referred to \cite{Milewski1999} for more details). In numerical experiments,  $1024\times256$ Fourier modes are used along with the propagating and transverse directions, respectively, and the computations are de-aliased by a doubling of Fourier modes. We take two typical plane solitary waves: $||\eta||_\infty=0.1894$, $c=0.9810$ (small amplitude) and $||\eta||_\infty=0.8069$, $c = 0.9490$ (large amplitude), and perturb them in the transverse direction using a harmonic function. Specifically, we set
\[
q(x,y,0)=\left[1+0.05\cos\left(k_2y\right)\right]U(x)\,,
\]
where $U(x)$ is the exact plane solitary-wave solution and $k_2$ is the wavenumber of perturbation. For the small-amplitude case, the critical wavenumber $k_\mathrm{c}\approx0.079$, and thus we choose the perturbation wavenumber $k_2=0.086$ and $0.070$. For the large-amplitude case, since $k_\mathrm{c}\approx0.188$, we take $k_2=0.208$ and $0.168$. It is shown in Fig. \ref{fig:6} that in both cases, the first perturbation does not destabilize the solitary wave, whereas the second one does, supporting the results illustrated by Fig. \ref{fig:5}. The evolution of the instability shows a focusing behavior and eventually results in a depression lump with a radiated wave field. We should point out that the transverse instability of plane solitary waves implies that depression lumps are stable and appear to be attractors in the long-time dynamics of this system (see Fig. \ref{fig:7}).

\subsection{Collisions}
As discussed above, depression lumps are stable, so it is natural to inquire into the dynamics of these solutions. We consider two types of wave interactions: the head-on and overtaking collisions of a pair of free lumps. In all the collision experiments, initial data are constructed by first shifting the lump solutions and then adding them together.

\begin{figure*}[htbp]
	\subfigure[]{
		\includegraphics[width=0.45\linewidth]{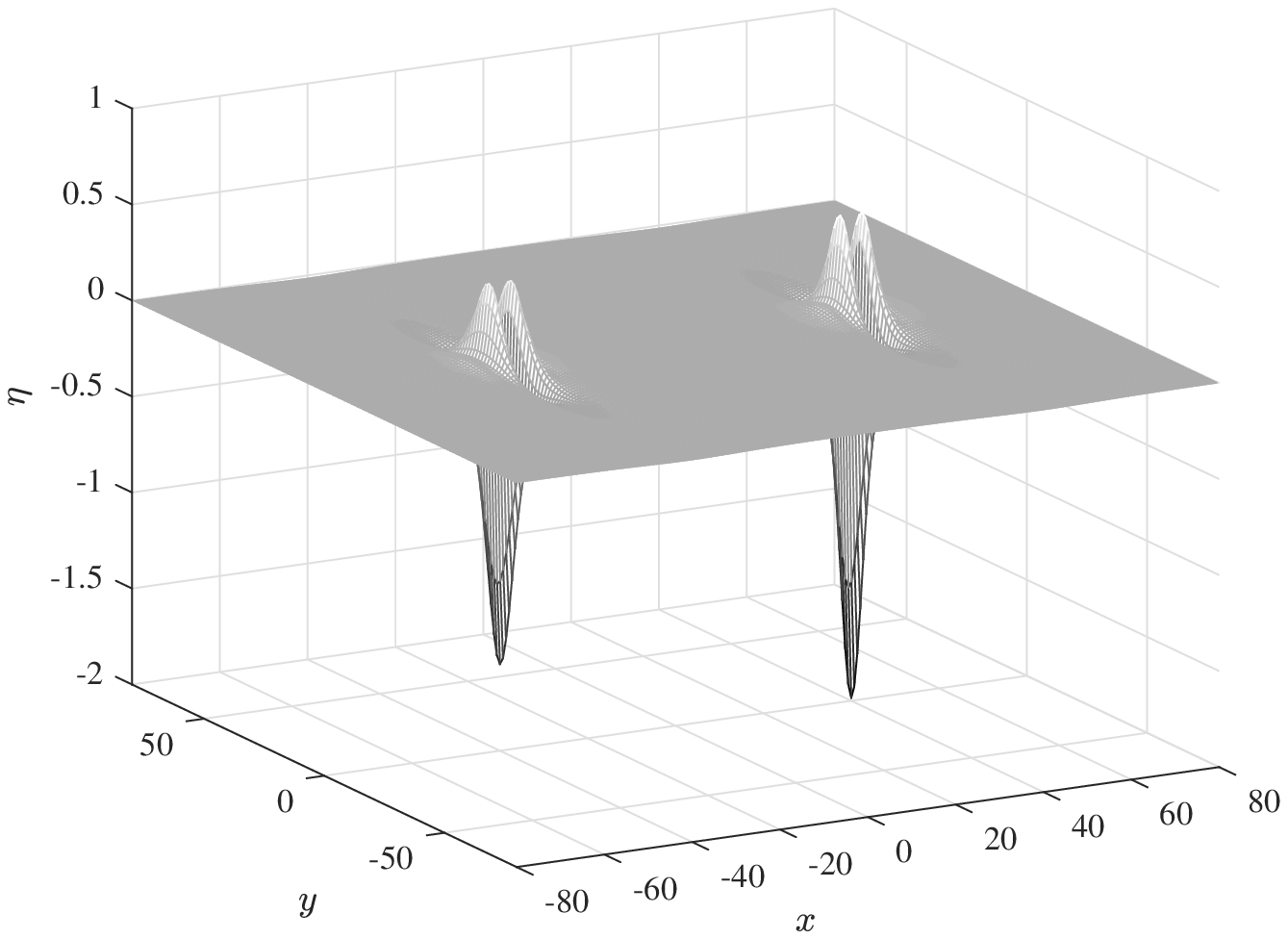}
		\label{fig:8a}}
	\subfigure[]{
		\includegraphics[width=0.45\linewidth]{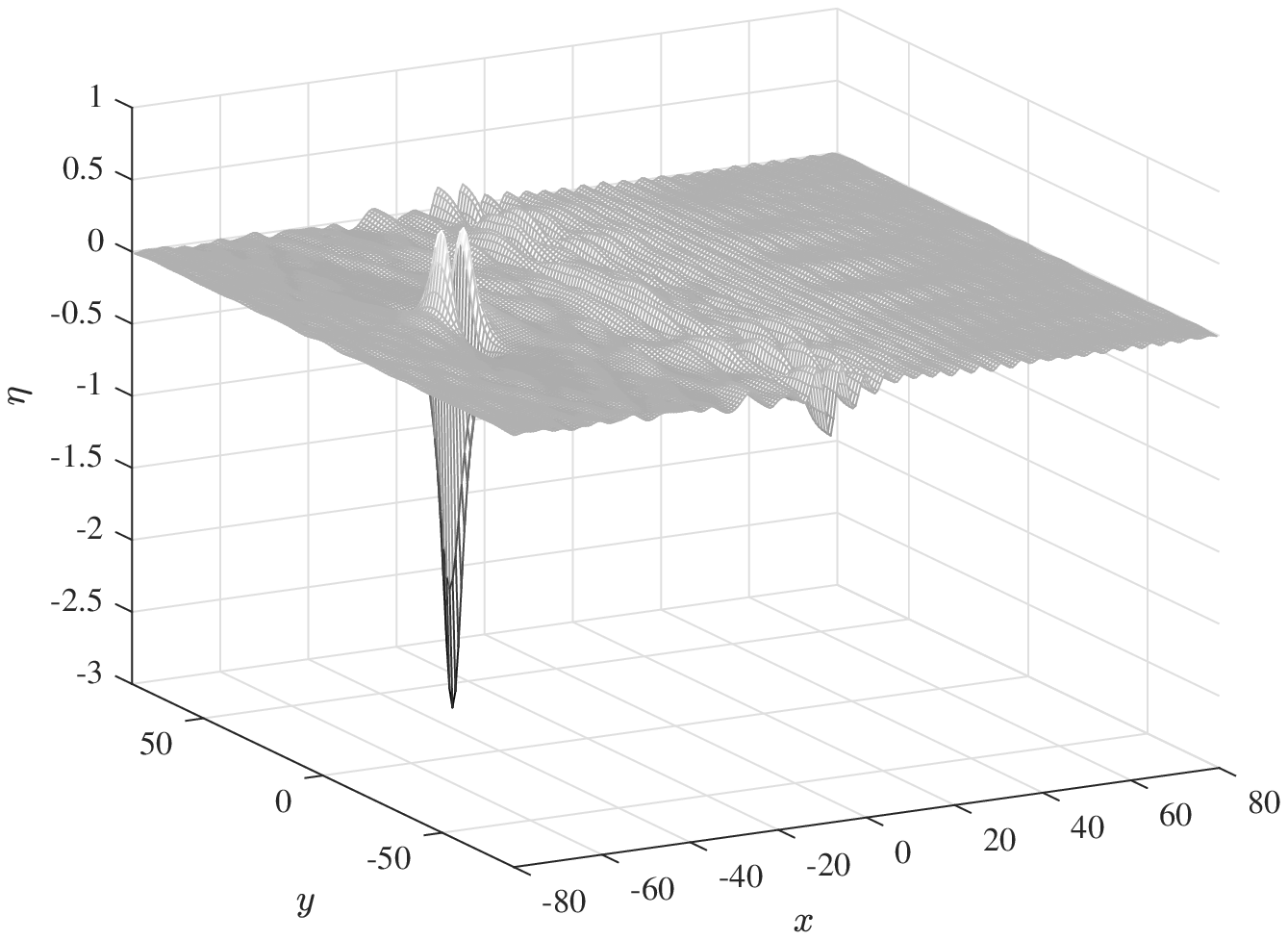}
		\label{fig:8b}}
	\caption{An overtaking collision between two lumps traveling in the positive $x$-direction. (a) Initially, a faster lump ($c=0.9476$, $||\eta||_\infty=1.5306$) is placed behind a slower one ($c=0.9353$, $||\eta||_\infty=1.9705$). (b) The wave profile after the collision ($t=5320$). The solution is shown in a frame of reference, moving at the speed of the left lump.}
	\label{fig:8}
\end{figure*}

There is no strongly nonlinear effect due to a short interaction time for a head-on collision between two depression lumps traveling along the x-axis in opposite directions. Hence, we do not show results of this type of collision in the form of figures. On the other hand,  \citet{Gao2016} showed two possibilities for overtaking collisions in the two-dimensional hydroelastic wave problem: both waves survive when they have a small initial difference in amplitude, and only one wave survives when the initial difference is relatively significant. Several numerical experiments are carried out in three dimensions by choosing lumps traveling in the same direction with different speeds. In all tested cases, only one lump can survive the overtaking collision regardless of the initial amplitude difference (see Fig. \ref{fig:8}). The amplitude of the resultant lump after the collision is greater than that of the original two, and dispersion oscillations are generated during lump interactions. 

\begin{figure*}[htbp]
	\centering
	\subfigure[]{
		\includegraphics[width=0.21\linewidth]{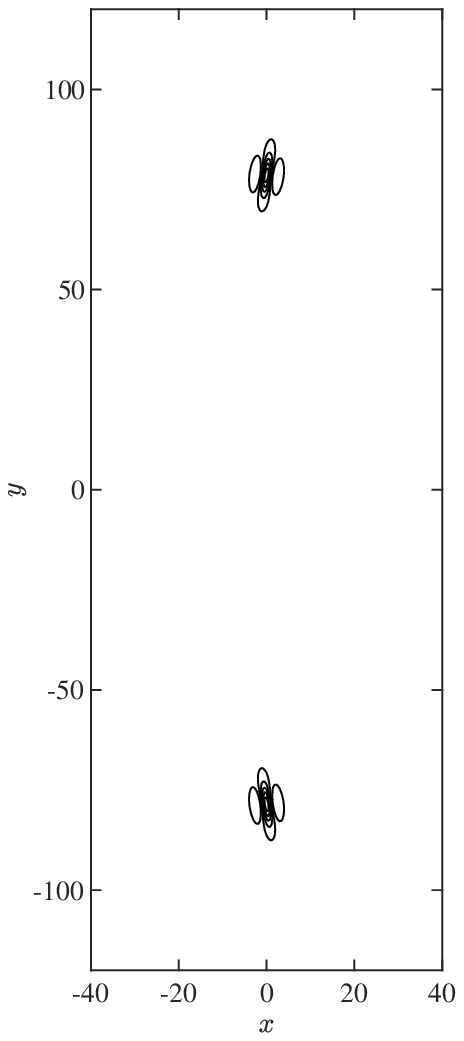}
		\label{fig:9a}}
	\subfigure[]{
		\includegraphics[width=0.21\linewidth]{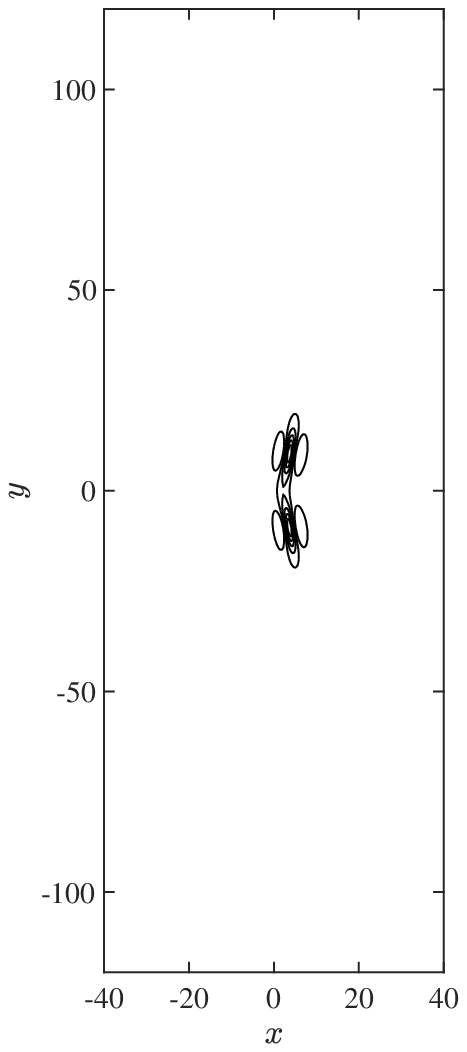}
		\label{fig:9b}}
	\subfigure[]{
		\includegraphics[width=0.21\linewidth]{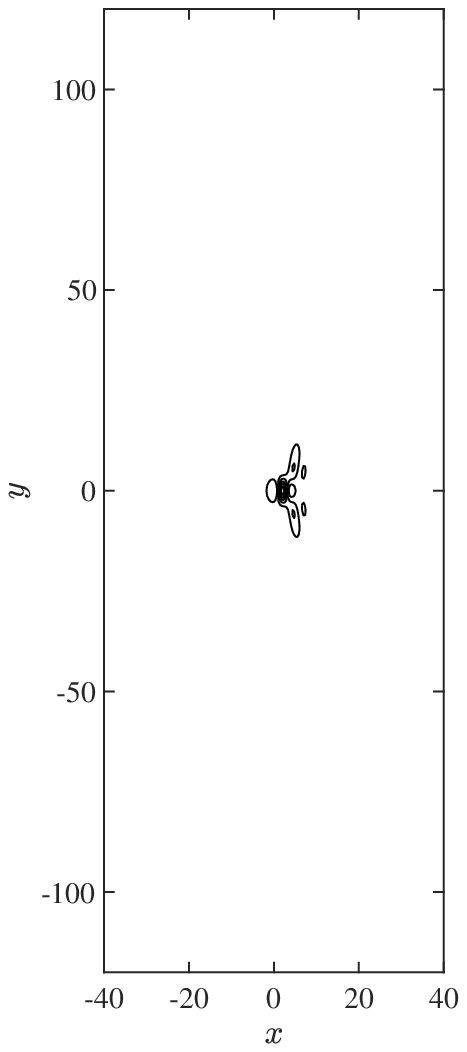}
		\label{fig:9c}}
	\subfigure[]{
		\includegraphics[width=0.21\linewidth]{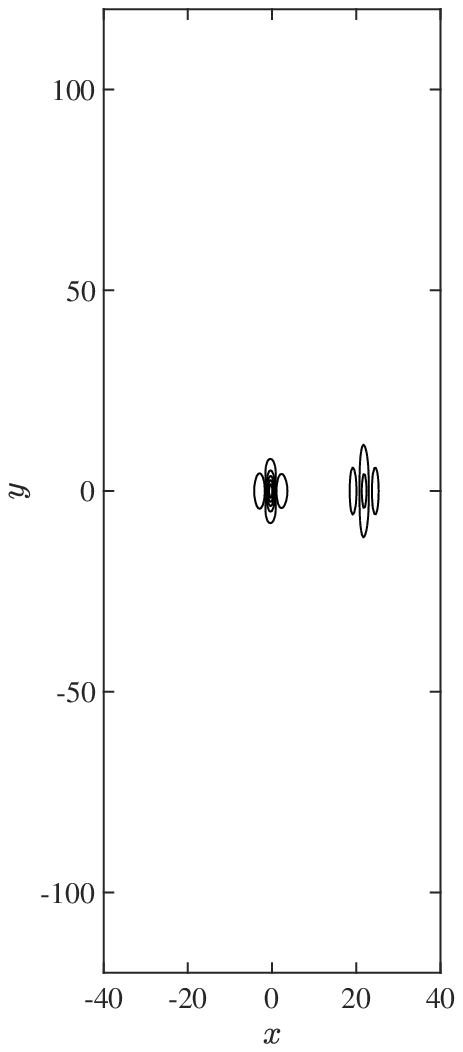}
		\label{fig:9d}}
	\caption{Contour plots of an oblique collision between two lumps of the same amplitude ($||\eta||_\infty=4.0992$). The computational domain is $60\pi\times180\pi$ with $256\times512$ discretization points. We choose the time step $\Delta{t}=0.05$ and the shallowness parameter $\mu=0.3$.}
	\label{fig:9}
\end{figure*}

It is noted the Benney-Luke-type equation is isotropic, and lumps can travel in any horizontal direction, which stimulates us to explore oblique collisions between lumps further. The oblique interaction of two lumps in the Benney-Luke-type equation is simulated in a $60\pi\times180\pi$ computational domain with a $256\times512$ grid. Two lumps of the same amplitude are initially placed on two sides of the line $y=0$ with mirror symmetry and travel towards the axis of symmetry at the incidence angle $\alpha=\arctan\tfrac18$. Fig. \ref{fig:9} shows the contour snapshots of the solution at different moments, and a nonlinear interaction with the generation of a radiation field can be observed. Our computation shows that the oblique collision is obviously inelastic since lumps undergo a significant phase shift: two waves first merge and then break up into two lumps moving along the central line in tandem. 

\subsection{Generation of lumps by a moving load}
Elastic ice covers are widely used as roadways to support transport systems in cold regions. One of the most important aspects of hydroelastic waves is to understand wave responses to moving loads on ice cover, which can be explored in the shallow water regime based on the forced Benney-Luke-type equation \eqref{BL2}. In computations of the forced problem, we use the pressure distribution 
\begin{equation*}
p(x, y, t)=-\text{sech}^2\left[0.3^2(x-Ut)^2+0.15^2y^2\right]\,, 
\end{equation*}
and results are qualitatively similar for other locally confined distributions. 

\begin{figure}[htbp]
	\centering
	\includegraphics[width=\linewidth]{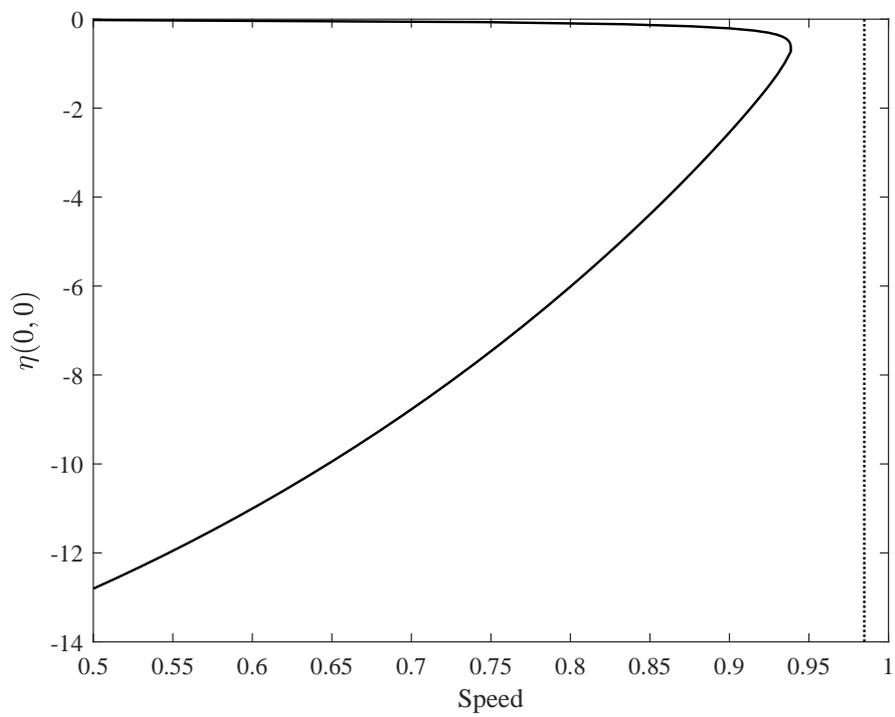}
	\caption{Bifurcation diagram of permanent wave solutions to the forced Benney-Luke-type equation. The vertical dotted line shows the phase speed minimum of the free problem.}
	\label{fig:10}
\end{figure}

Permanent wave solutions to the forced equation are solved using Newton's method, and the whole solution branch can be obtained through a numerical continuation method. It is shown in Fig. \ref{fig:10} that there is an interval of load speed, $c^*<U<c_{\mathrm{min}}$, where no steady-state responses exist. Here $c^*$ is denoted as the largest forcing speed for which there exists a permanent wave solution (for example, $c^*\approx0.9348$ for $\mu=0.3$) and $c_{\mathrm{min}}$ represents the minimum of the phase speed (i.e., the bifurcation point of free solitary waves). The existence of this transcritical region of forcing speed naturally leads us to explore the dynamic transient responses, which are frequently associated with the phenomenon of shedding of solitary waves. That is because when the forcing speed is in the transcritical regime, there is neither a steady forced solution nor linear mechanism to radiate energy away, and a possible way to resolve the unlimited accumulation of energy is to release it in the form of solitary waves. Examples are \citet{Wu1987} who showed numerically the periodic generation of upstream-propagating pure gravity solitary waves under a forcing moving at a transcritical speed, \citet{Zhu1986} who observed the same phenomenon in an experiment for a stratified fluid, and \citet{Berger2000} who obtained similar results for three-dimensional gravity-capillary flows in the shallow water regime.

\begin{figure}[htbp]
\centering
\subfigure[]{
	\includegraphics[width=\linewidth]{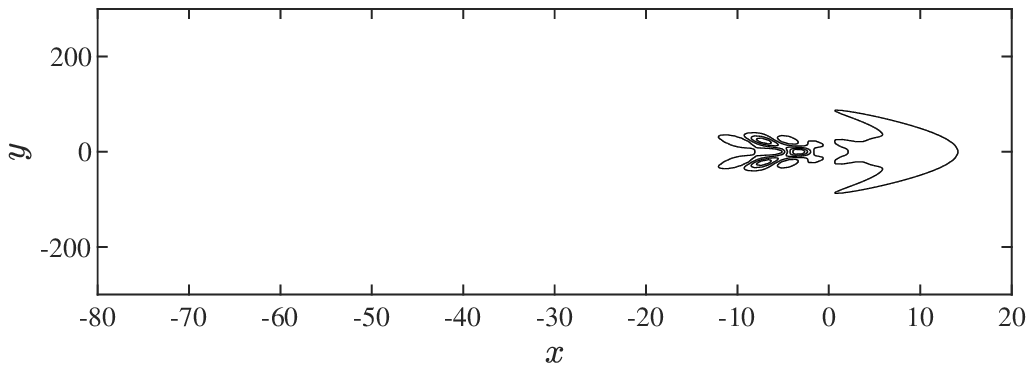}
	\label{fig:11a}}
\subfigure[]{
	\includegraphics[width=\linewidth]{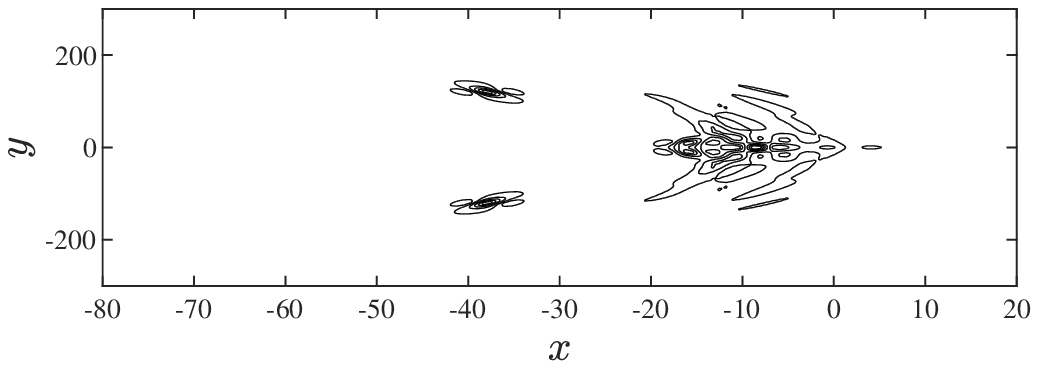}
	\label{fig:11b}}
\subfigure[]{
	\includegraphics[width=\linewidth]{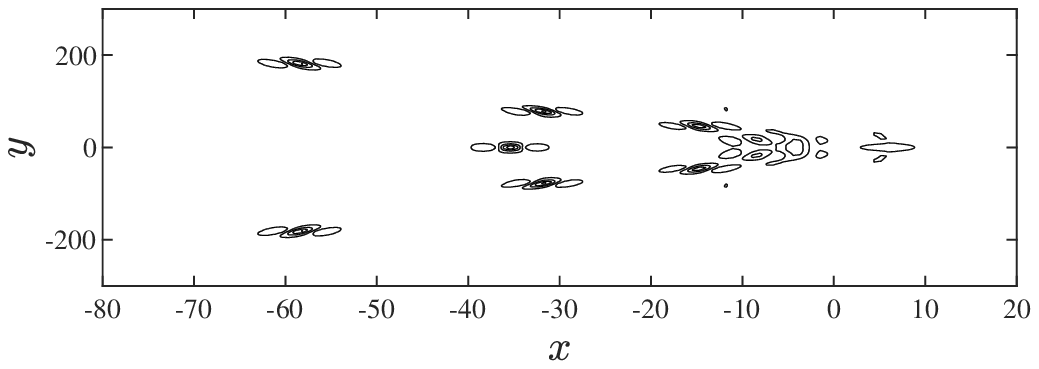}
	\label{fig:11c}}
\caption{Contour plots of ice responses to a load moving at $U=0.98$ for $t=450$, $t=1500$, and $t=2100$ from top to bottom.}
\label{fig:11}
\end{figure}

\begin{figure}[htbp]
\centering
\subfigure[]{
	\includegraphics[width=\linewidth]{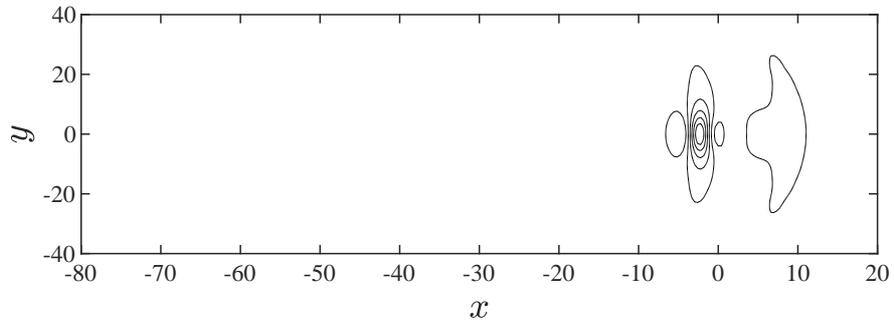}
	\label{fig:12a}}
\subfigure[]{
	\includegraphics[width=\linewidth]{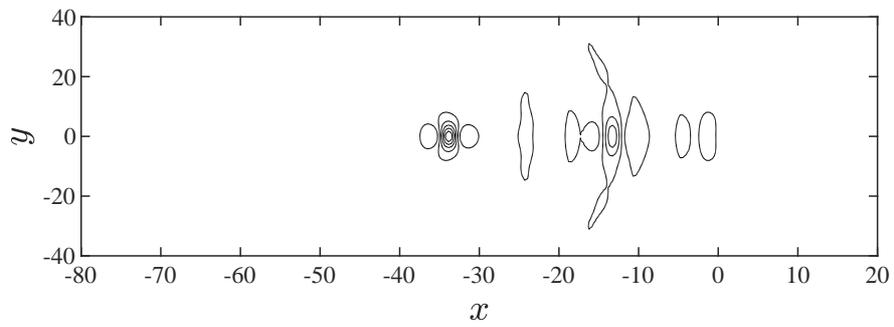}
	\label{fig:12b}}
\subfigure[]{
	\includegraphics[width=\linewidth]{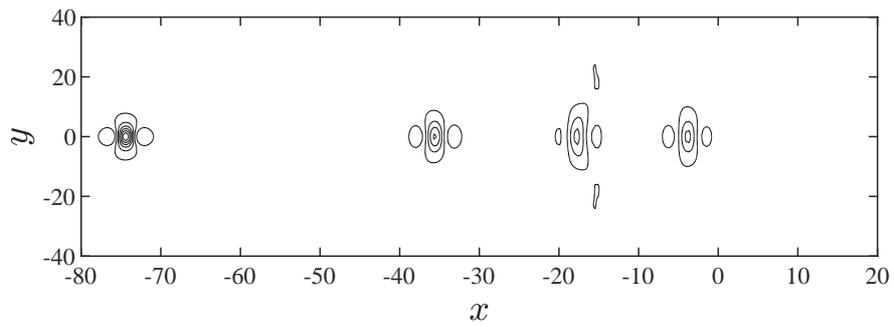}
	\label{fig:12c}}
\caption{Contour plots of ice responses to a load moving at $U=0.95$ for $t = 140$, $t = 990$, and $t=1350$ from top to bottom.}
\label{fig:12}
\end{figure}

\begin{figure*}[htbp]
	\subfigure[]{
		\includegraphics[width=0.44\linewidth]{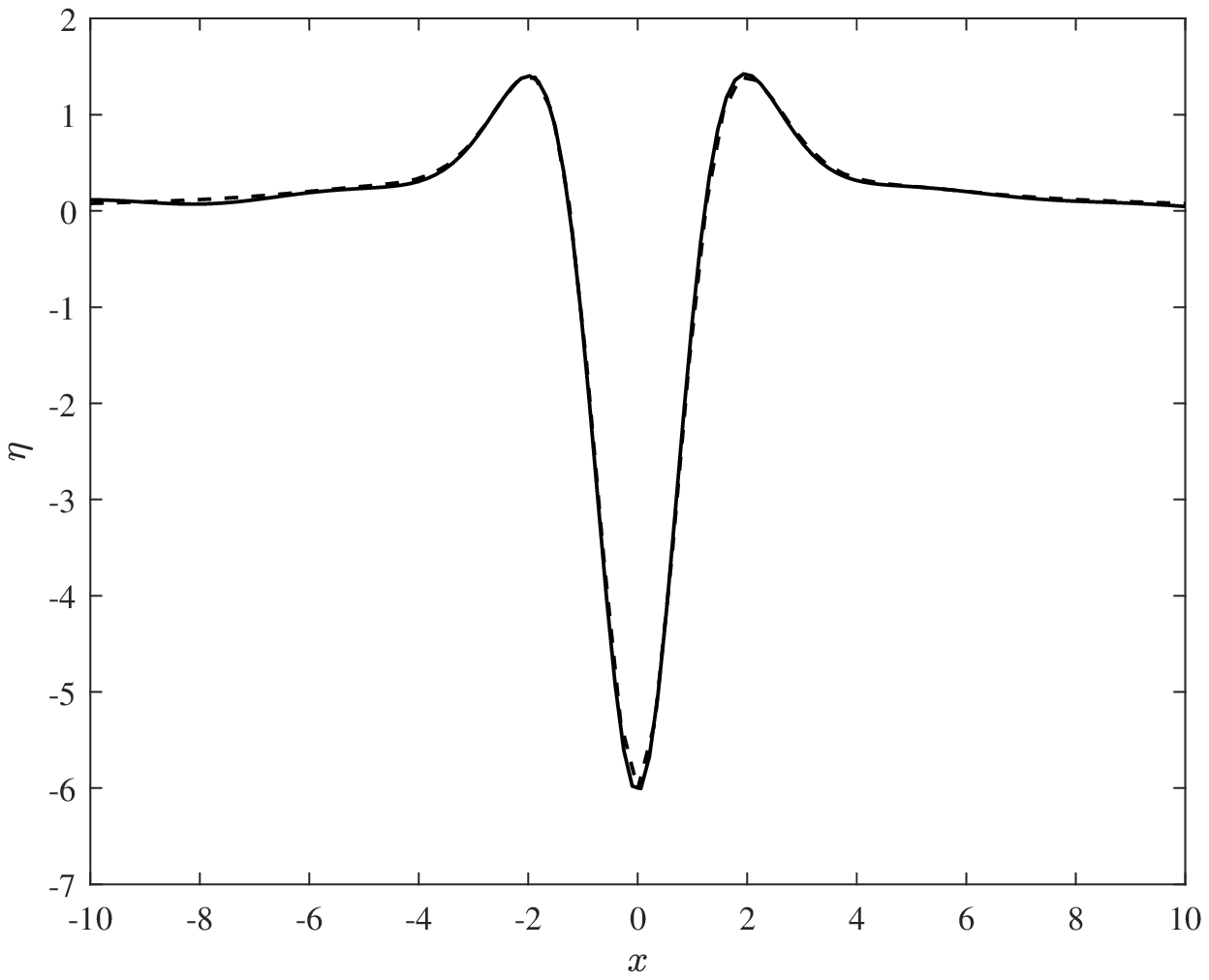}
		\label{fig:13a}}
	\subfigure[]{
		\includegraphics[width=0.44\linewidth]{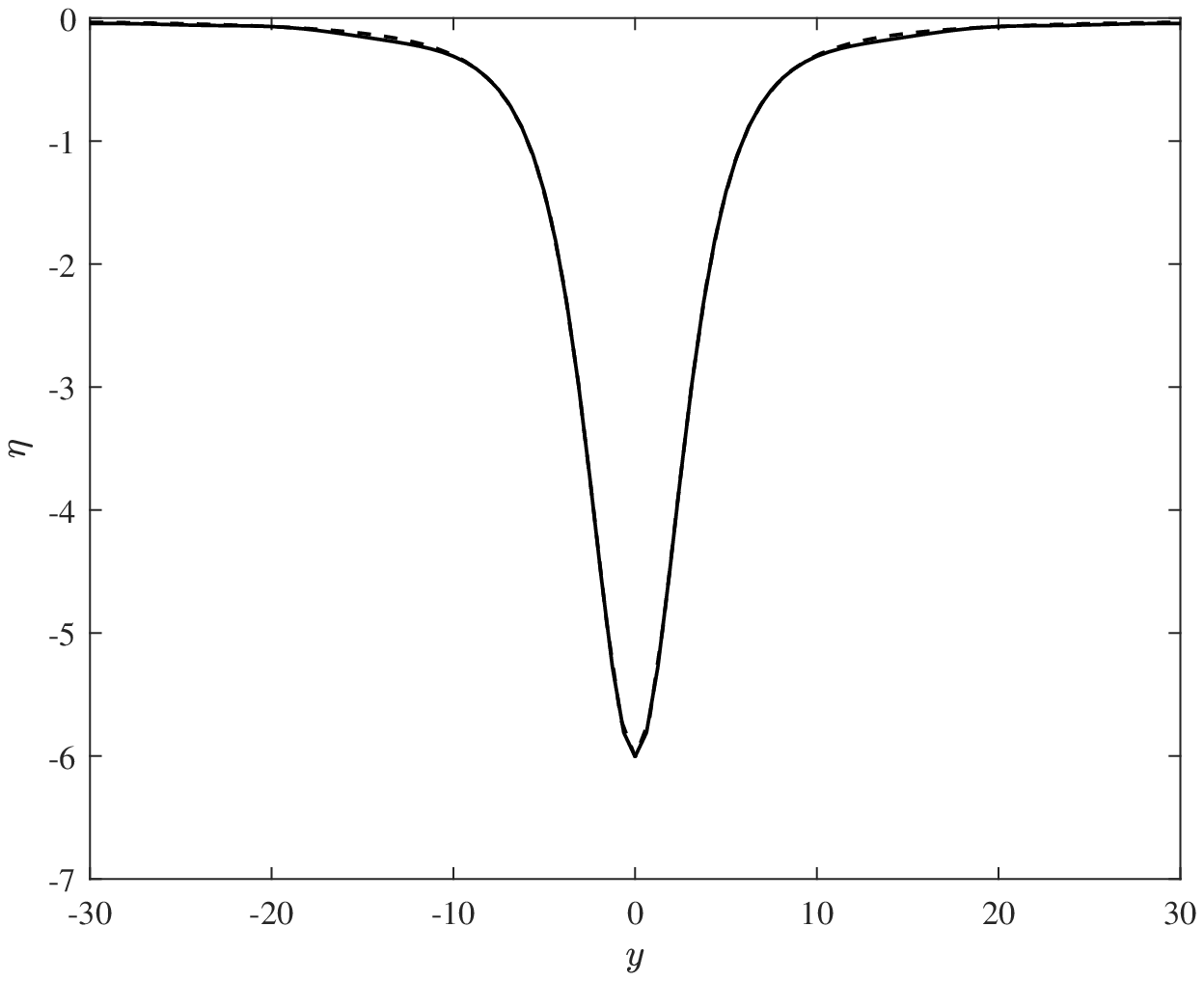}
		\label{fig:13b}}
	\caption{Comparisons of the generated lump shown at the leftmost edge of Fig. \ref{fig:12c} (solid line) to the exact steady solution (dashed line): (a) for $x$-cross-section and (b) for $y$-across-section.}
	\label{fig:13}
\end{figure*}

Next, we examine the transient solutions of the forced hydroelastic wave problem in the transcritical regime. Two representative cases of ice responses are shown in Fig. \ref{fig:11} and Fig. \ref{fig:12}. For $U=0.98$, a time-periodic shedding of lumps occurs, and these lumps propagate along three lines behind the forcing (see Fig. \ref{fig:11}). We also check the cases of $U=0.97$ and $U=0.96$, and the same phenomenon can be observed. It is found that if the forcing is moving at speed less than $0.958$, lumps emerge merely along the negative $x$-axis, and a typical example is shown in Fig. \ref{fig:12} for the forcing moving at $U = 0.95$. It seems plausible that the shedding routes of lumps vary with the forcing speed. For speeds closer to $c_{\min}$, say $U=0.98$, more energy is accumulated, giving rise to multiple lumps released almost simultaneously. When the load speeds are a bit away from the phase speed minimum, less energy builds up, and only one lump forms at a time. Finally, Fig. \ref{fig:13} demonstrates the comparisons of $x$- and $y$-cross-sections between the resultant lump located at the leftmost edge of Fig. \ref{fig:12c} and the exact solution, which show a remarkable agreement and validate our numerical codes as well. 


\section{Conclusion}
In this paper, we are concerned with hydroelastic solitary waves on the surface of a three-dimensional fluid. The long-wave approximation is made, and a Benney-Luke-type equation possessing a Hamiltonian structure has been derived via the explicit non-local formulation of water waves introduced by \citet{Ablowitz2006}. For the newly developed model, the associated BRDS system governing the envelope dynamics of a quasi-monochromatic plane wave has been established and numerically solved to help understand the existence, asymptotic behavior, and stability properties of solitary waves in the Benney-Luke-type equation.
 
For the unforced problem, three types of solitary waves, which can be predicted by a one-parameter family of solutions in the associated BRDS system, have been numerically computed in the Benney-Luke-type equation. We then discussed the asymptotic behavior of lumps based on the envelope equation, and it was shown that these lumps featured algebraically decaying tails in the far field. The transverse instability and subsequent evolution of plane solitary waves have been thoroughly explored. We proved that plane solitary waves were unstable subject to long-wavelength transverse perturbations and specified the critical perturbation wavenumber for the onset of instability, which coincided with the dimension-breaking bifurcation point. Numerical experiments have been conducted to verify these results, which showed that the transverse instability of plane solitary waves resulted in the emergence of depression lumps. Besides, interactions between lumps, including the head-on, overtaking, and oblique collisions, have also been investigated via numerical time integrations.

When a constant-moving load forces the problem, it has been found that there is a transcritical range of forcing speed where no permanent wave solutions can exist. Several numerical experiments have been carried out to study the transient responses to the forcing in the transcritical regime. The periodic shedding of lumps behind the forcing has been observed, indicating another generation mechanism of lumps. The shedding routes of lumps are qualitatively different for $U=0.95$ and $U=0.98$, which should be relevant to the amount of accumulated energy during the period. 

Finally, we should point out that our model can be extended by including the bottom topography and viscoelasticity of the plate. Therefore the Bernoulli equation can be rewritten as 
\begin{equation*}
\phi_t+\frac12\left(|\nabla\phi|^2+\phi_z^2\right)+g\eta+\frac{D}{\rho_w}\Delta^2\eta+\frac{\rho_id}{\rho_w}\eta_{tt}+\frac{\nu}{\rho_w}\eta_t=p\,.
\end{equation*}
The impermeability boundary condition at the bottom becomes
\begin{equation*}
\phi_z-\nabla b\cdot\nabla\phi=0\,,
\end{equation*}
where the factor $\nu>0$ in the damping term is assumed to be constant, and $z=-h+b(x,y)$ is the bottom topography with $||b||_\infty=O\left(||\eta||_\infty\right)$. Following the same derivation procedure presented in \S2, one can obtain
\begin{equation*}
q_{tt}-\Delta q-\left(\frac{\mu^2}{3}+\gamma\right)\Delta^2q-\delta\Delta^3q+\epsilon\left(\partial_t|\nabla q|^2+q_t\Delta q\right)-\widetilde{\nu}\Delta q_t+\epsilon\nabla\cdot b\nabla q=\epsilon p_t\,,
\end{equation*}
where the dimensionless damping coefficient $\widetilde{\nu}$ is assumed to be $O(\epsilon)$.  

\section*{Acknowledgement}
This work was supported by the Key Research Program of Frontier Sciences of CAS (No.
QYZDB-SSW-SYS015) and the Strategic Priority Research Program of CAS (No. XDB22040203). 

\bibliographystyle{unsrtnat}
\bibliography{paper}
\end{document}